\documentclass[12pt,a4wide]{article}

\usepackage{epsf}
\usepackage{longtable}
\usepackage{amsmath}
\usepackage{amssymb}
\usepackage{epubtk}
\usepackage{fancybox}
\usepackage{graphicx}

\newcommand{\gym}{g_{\rm YM}}
\newcommand{\gs}{g_S}
\newcommand{\Nfour}{${\cal N}=4$ }

\newcommand{\AdSS}{$AdS_5\times S^5$ }
\newcommand{\cO}{{\cal O}}
\newcommand{\Tr}{\mbox{Tr}}
\newcommand{\ft}[2]{{\textstyle \frac{#1}{#2}}}
\newcommand{\eqn}[1]{(\ref{#1})}
\newcommand{\be}{\begin{equation}}
\newcommand{\ee}{\end{equation}}
\newcommand\bea{\begin{eqnarray}}
\newcommand\eea{\end{eqnarray}}
\newcommand\beal{\begin{align}}
\newcommand\eeal{\end{align}}
\newcommand\nn\nonumber
\newcommand{\up}{\uparrow}
\newcommand{\down}{\downarrow}
\newcommand{\Zint}{\mathbb{Z}}      
\newcommand{\Real}{\mathbb{R}}            
\newcommand{\eins}{{\bf 1}}      

\newlength{\intwidth}
\DeclareRobustCommand{\fpint}[2]
  {\mathop{%
     \text{%
       \settowidth{\intwidth}{$\int$}%
       \makebox[0pt][l]{\makebox[\intwidth]{$-$}}%
       $\int_{#1}^{#2}$}}}

\begin{document}

\title{Spinning strings and integrable spin chains in the AdS/CFT correspondence}

\author{%
Jan Plefka\\
Max-Planck-Institut f\"ur Gravitationsphysik\\
(Albert-Einstein-Institut)\\
Am M\"uhlenberg 1, 14476 Potsdam, Germany\\
{\tt jan.plefka@aei.mpg.de}}

\date{{\small AEI-2005-124, hep-th/0507136}}
\maketitle

\begin{abstract}
In this introductory review we discuss dynamical tests of the \AdSS string/\Nfour super
Yang-Mills duality. After a brief introduction to AdS/CFT we argue that semiclassical string energies
yield information on the quantum spectrum of the string in the limit
of large angular momenta on the $S^5$.
The energies of the folded and circular spinning string solutions
rotating on a $S^3$ within the $S^5$ are derived,
which yield all loop predictions for the dual gauge
theory scaling dimensions. These follow from the eigenvalues of the dilatation operator of 
\Nfour super Yang-Mills in a minimal $SU(2)$ subsector and we display its reformulation in terms of a
Heisenberg $s=1/2$ spin chain along with the coordinate Bethe ansatz for its explicit diagonalization. 
In order to make
contact to the spinning string energies we then study the thermodynamic limit of the one-loop gauge
theory Bethe equations and demonstrate the matching with the folded and closed string result at this
loop order. Finally the known gauge theory results at higher-loop orders are reviewed and the 
associated long-range spin chain Bethe ansatz is introduced, leading to an asymptotic all-loop
conjecture for the gauge theory Bethe equations. This uncovers discrepancies at the three-loop
order between gauge theory scaling  dimensions and string theory energies and the implications
of this are discussed. Along the way we comment on further developments and generalizations
of the subject and point to the relevant literature. 
\end{abstract}

\epubtkKeywords{String Theory, Supersymmetric Yang-Mills, AdS/CFT Correspondence}

\newpage


\section{Introduction}
\label{section:introduction}

String theory was initially discovered in  an attempt to describe the physics
of the strong interactions prior to the advent of gauge field theories and
QCD. Today it has matured to a very promising candidate for a unified 
quantum theory of gravity and all the other  forces of nature.
In this interpretation gauge fields arise as the low energy
excitations of fundamental open strings and are therefore
derived, non-fundamental objects, just as the theory of gravity itself. 
Ironically though, advances in our understanding
of non-perturbative string theory and of D-branes has led to a resurrection of gauge fields
as fundamental objects. Namely it is now generally believed that
string theory in suitable space-time backgrounds can have a dual, holographic description in 
terms of gauge field theories and thus the question
of which of the two is the fundamental one becomes redundant. This
belief builds on a remarkable proposal due to Maldacena \cite{hep-th/9711200} known
as the Anti-de-Sitter/Conformal Field Theory (AdS/CFT) 
correspondence (for reviews see \cite{hep-th/9905111,hep-th/0201253}).

The initial idea of a string/gauge duality is due to 't Hooft \cite{'tHooft:1973jz}, who 
realized that the perturbative expansion of $SU(N)$ gauge field
theory in the large $N$ limit can be reinterpreted as a genus expansion of
discretized two dimensional surfaces built from the field theory Feynman diagrams.
Here $1/N$ counts the genus of the Feynman diagram, while the 't Hooft coupling
$\lambda:=\gym^2\, N$ (with $\gym$ denoting the gauge theory coupling constant)
enumerates quantum loops. 
E.g.~the genus expansion of the free energy $F$ of a  $SU(N)$ gauge theory in the 't
Hooft limit ($N\to\infty$ with $\lambda$ fixed) takes the pictorial form
\begin{equation}
F=N^2\,\raisebox{-.3cm}{\epsfysize=1cm\epsfbox{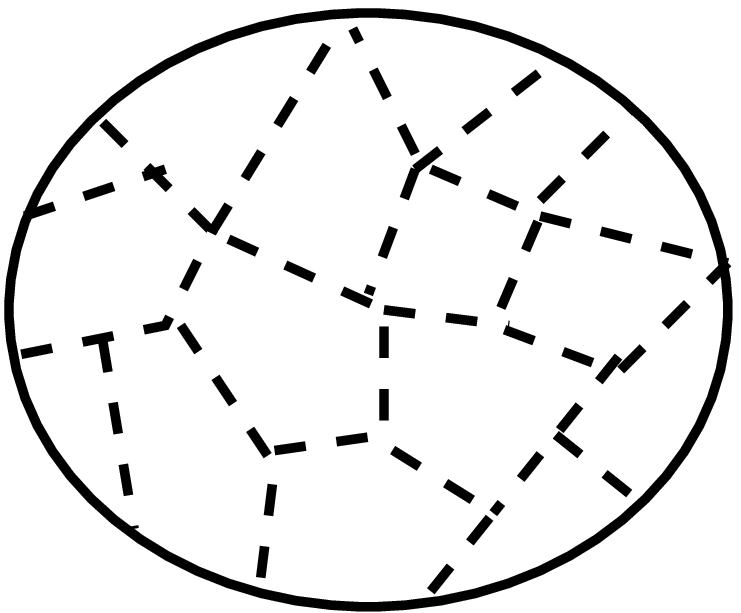}}
+ {1}\,  \,\raisebox{-.3cm}{\epsfysize=1cm\epsfbox{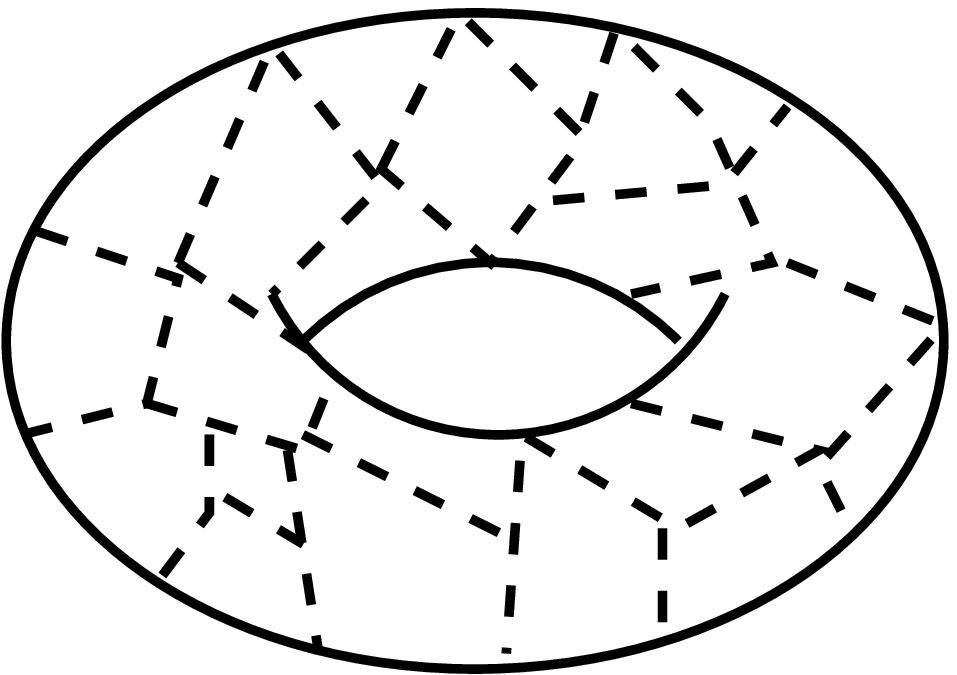}}
+ {\frac{1}{N^2}}\,  \,\raisebox{-.3cm}{\epsfysize=1cm\epsfbox{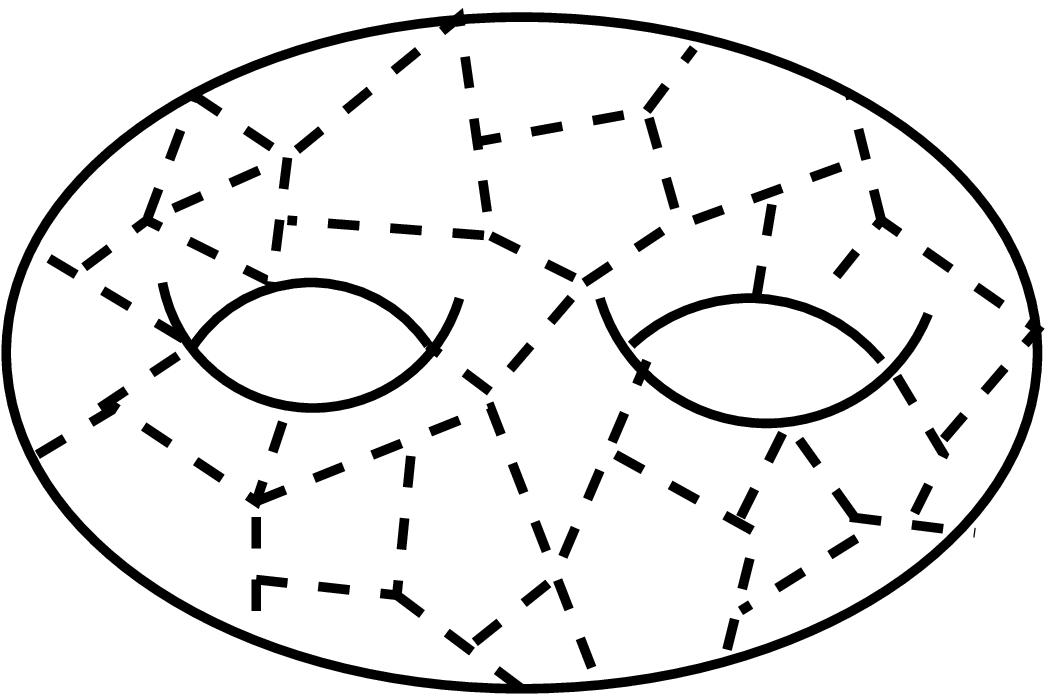}}
+\ldots 
 =\sum_{g=0}^\infty \frac{1}{{N^{2g-2}}}\, \sum_{l=0}^\infty c_{g,l}\, 
{\lambda}^l
\end{equation}
with suitable coefficients $c_{g,l}$ denoting the contributions at genus $g$ and loop
order $l$. Obviously this $1/N$ expansion resembles the perturbative expansion of a 
string theory in the string coupling constant $g_S$.

The AdS/CFT correspondence is the first concrete realization of this
idea for four dimensional gauge theories.
In its purest form -- which shall also be the setting we will be
interested in -- it identifies the `fundamental'
type IIB superstring in a ten dimensional
 anti-de-Sitter cross sphere ($AdS_5\times S^5$) space-time background with the 
maximally supersymmetric Yang-Mills theory with gauge group $SU(N)$
(\Nfour SYM) in four dimensions. The \Nfour super Yang-Mills model is a
quantum conformal field theory, as its $\beta$-function vanishes exactly. The string model 
is controlled by two parameters: The string coupling constant $\gs$
and the `effective' string tension
$R^2/\alpha'$ where $R$ is the common radius of the 
$AdS_5$ and $S_5$ geometries. The gauge theory, on the other hand, is parametrized by the rank
$N$ of the gauge group and the coupling constant $\gym$, or equivalently the 't Hooft coupling
$\lambda:=\gym^2\, N$. According to the AdS/CFT proposal, these two sets of parameters are to be
identified as
\begin{equation}
\label{eq:AdS/CFT rels}
\frac{4\pi\lambda}{N}=\gs \qquad \sqrt{\lambda} = \frac{R^2}{\alpha'} \, .
\end{equation}
We see that in the AdS/CFT proposal the string coupling constant is
not simply given by $1/N$, but comes with a linear factor in
$\lambda$. This, however, does not alter the genus expansion and its
interpretation in form of string worldsheets.

The equations \eqn{eq:AdS/CFT rels} relate the coupling constants but there is
also a dictionary between the excitations of the two theories.
The correspondence identifies the energy eigenstates of the \AdSS
string, 
which we denote schematically as $|\cO_A\rangle$ with $A$ being a multi-index,  
with (suitable) composite gauge
theory operators of the form 
$\cO_{A}=\Tr(\phi_{i_1}\, \phi_{i_2}\ldots\phi_{i_n})$ where
$(\phi_{i})_{ab}$ are the
elementary fields of \Nfour SYM (and their covariant derivatives) in the adjoint
representation of $SU(N)$, i.e.~$N\times N$ hermitian matrices . 
The energy eigenvalue $E$ of  a string state with respect to time in
global coordinates 
is conjectured to be equal to the scaling dimension $\Delta$ of the dual gauge theory operator,
which in turn is determined from the two point function of the conformal field theory
\footnote{For simplicity we take ${\cal O}_A(x)$ to be a scalar
  operator here.}
\begin{equation}
\label{eq:DvsE}
\langle \cO_A(x)\, \cO_B(y) \rangle = \frac{M\,
  \delta_{A,B}}{(x-y)^{2\, \Delta_A(\lambda,\ft 1 N)}}
\quad\Leftrightarrow\quad
{\cal H}_{\rm String}\, |\cO_A\rangle = E_A(\ft{R^2}{\alpha'},\gs)\,
|\cO_A\rangle 
\end{equation}
with $\Delta(\lambda,\ft 1 N) \stackrel{!}{=} E(\ft{R^2}{\alpha'},\gs)$.

A zeroth order test of the conjecture is then the agreement of the
underlying symmetry supergroup $PSU(2,2|4)$ of the
two theories, which furnishes the representations under which $\cO_A(x)$
and $|\cO_A\rangle$  transform. This then yields 
a hint on how one could set up an explicit string state/gauge operator dictionary. 

Clearly there is little hope of determining either the all genus 
(all orders in $\gs$) string spectrum or the complete $1/N$ dependence of the gauge theory scaling
dimensions $\Delta$. But the identification of the planar gauge theory with the free ($\gs=0$)
string seems feasible and fascinating: Free \AdSS string theory
should give the exact all loop
gauge theory scaling dimensions in the large $N$ limit! Unfortunately though, 
our knowledge of the string spectrum in 
curved backgrounds, even in such a highly symmetric one as \AdSS,
remains scarce.  Therefore until very recently
investigations on the string side of the correspondence were limited to the
domain of the low energy effective field
theory description of \AdSS strings in terms of type IIB supergravity. This, however, is necessarily
limited to weakly curved geometries in string units, i.e.~to the domain of $\sqrt{\lambda}\gg 1$
by virtue of \eqn{eq:AdS/CFT rels}. On the gauge theory side one has control only
in the perturbative regime where $\lambda\ll 1$, which is perfectly incompatible
with the accessible supergravity regime $\sqrt{\lambda}\gg 1$. Hence one is facing 
a strong/weak coupling 
duality, in which strongly coupled gauge fields are described by {\sl classical} supergravity and
weakly coupled gauge fields correspond to strings propagating in a highly curved background
geometry. This insight is certainly fascinating, but at the same time strongly hinders any dynamical tests
(or even a proof) of the AdS/CFT conjecture in regimes which are not protected by the
large amount of symmetry in the problem.

This situation has profoundly changed since 2002 by performing studies of the correspondence in 
novel limits where quantum numbers (such as spins or angular momenta in the geometric \AdSS language) 
become large in a correlated fashion as $N\to\infty$. 
This was initiated in the work of Berenstein, Maldacena
and Nastase \cite{hep-th/0202021} who considered the quantum fluctuation expansion of the
string around a degenerated point-like configuration, 
corresponding to a particle rotating with a large angular momentum $J$ on a
great circle of the $S^5$ space. In the limit of $J\to\infty$ with $J^2/N$ held fixed (the `BMN limit')
the geometry seen by the fastly moving particle is a
gravitational plane-wave, which allows for an exact quantization of the
free string in the light-cone-gauge \cite{hep-th/0112044,hep-th/0202109}. 
The resulting string spectrum leads to a formidable prediction
for the all loop scaling dimensions of the dual gauge theory operators in the corresponding
limit, i.e.~the famous formula $\Delta_n= J +2\, \sqrt{1+\frac{\lambda\,n^2}{J^2}}$ for the simplest
two string oscillator mode excitation.
The key point here is the emergence of the effective gauge theory 
loop counting parameter $\lambda/J^2$ in the BMN limit.
These scaling dimensions have by now been firmly reproduced up to the three loop order
in gauge theory \cite{hep-th/0303060,hep-th/0310252,hep-th/0409009}. 
This has  also led to important structural information for higher (or all loop)
attempts in gauge theory, which maximally employ the uncovered  
integrable structures to be discussed below. 
Moreover, 
the plane wave string theory/\Nfour SYM duality could be extended to the interacting string 
($g_S\neq 0$) respectively non-planar gauge theory regime providing us with the most concrete realization 
of a string/gauge duality to
date (for reviews see \cite{hep-th/0307027,hep-th/0307101,hep-th/0310119,hep-th/0401155}). 

In this review we shall discuss developments beyond the BMN plane-wave
correspondence which employ more
general sectors of large quantum numbers in the AdS/CFT duality. The key point from the
string perspective is that such a limit can make the semiclassical (in the plane-wave case) or even
classical (in the `spinning string' case) computation of the string
energies also quantum exact \cite{hep-th/0304255,hep-th/0306130}, 
i.e. higher $\sigma$-model
loops are suppressed by inverse powers of the total angular momentum
$J$ on the five sphere\footnote{This suppression
  only occurs if at least one angular momentum on the five sphere
  becomes large. If this is not the case, e.g.~for a spinnning string
  purely on the $AdS_5$ \cite{hep-th/0204051}, quantum corrections are not
  suppressed by inverse powers of the spin on $AdS_5$
  \cite{hep-th/0204226}.}. 
These considerations on the string
side then (arguably) yield all loop predictions
for the dual gauge theory. Additionally the perturbative gauge
theoretic studies at the first few orders in $\lambda$ led to
the discovery that the spectrum of scaling dimensions of the planar
gauge theory is identical to that of an integrable long-range spin
chain \cite{hep-th/0212208,hep-th/0303060,hep-th/0307042}. 
Consistently the $AdS_5\times S^5$ string
is a classically integrable model \cite{hep-th/0305116}, which has been heavily exploited in the
construction of spinning string solutions.

This review aims at a more elementary introduction to this very active area of research, which
in principle holds the promise of finding the exact quantum spectrum of the $AdS_5\times S^5$  
string or equivalently 
the all loop scaling dimensions of planar \Nfour super Yang-Mills. It is
intended as a first guide to the field for students and interested `newcomers'
 and points to the relevant literature for deeper studies.
We will discuss the simplest solutions of the $AdS_5\times S^5$ string corresponding to folded
and circular string configurations propagating in a $\Real\times
S^3$ subspace, with the $S^3$ lying within the $S^5$.
On the gauge theory side we will motivate the emergence of
the spin chain picture at the leading one loop order 
and discuss the emerging Heisenberg XXX${}_{1/2}$ model and its diagonalization
using the coordinate Bethe ansatz technique. This then enables us to perform a comparison 
between the classical string predictions in the limit of large angular momenta 
and the dual thermodynamic limit of the spin chain spectrum. Finally we turn to higher-loop
calculations in the gauge-theory and discuss conjectures for the all-loop form of the Bethe
equations, giving rise to a long-range interacting spin chain. Comparison with the obtained 
string results uncovers a discrepancy from loop order three onwards and the 
interpretation of this result is also discussed.

There already exist a number of more detailed reviews on spinning
strings, integrability and spin chains in the AdS/CFT correspondence:
Tseytlin's review \cite{hep-th/0311139} mostly focuses on the string side
the correspondence, whereas Beisert's Physics Report \cite{hep-th/0407277} concentrates 
primarily on the gauge side. See also Tseytlin's second review
\cite{hep-th/0409296}, on the so-called coherent-state
effective action approach, which we will not discuss in this review. 
Recommended is also the shorter review by
Zarembo \cite{hep-th/0411191} on the $SU(2)$ respectively $\Real\times S^3$
subsector, discussing the integrable structure appearing on the classical string -- not
covered in this review. For a detailed account of the near plane-wave superstring, its
quantum spectroscopy and integrability structures see Swanson's thesis \cite{hep-th/0505028}.


\section{The setup}
\label{section:setup}

\begin{figure}[t]
\label{fig:one}
\begin{center}
\includegraphics[width=2cm]{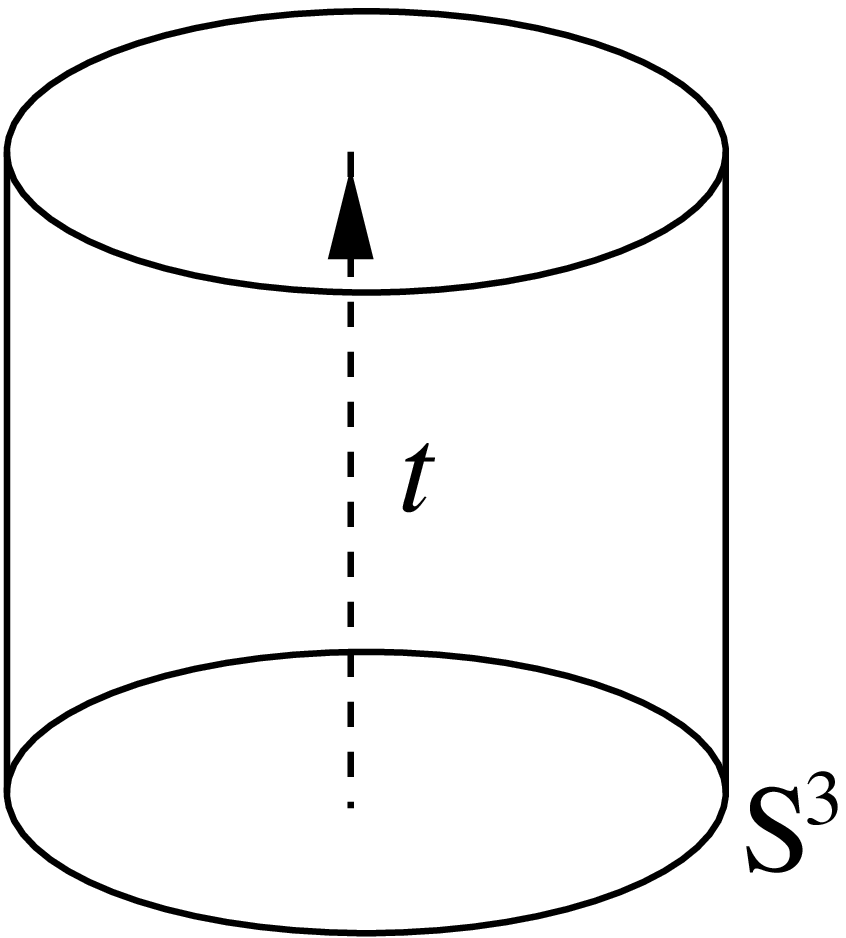}
\quad \raisebox{0.8cm}{$\times$} \quad
\includegraphics[width=2cm]{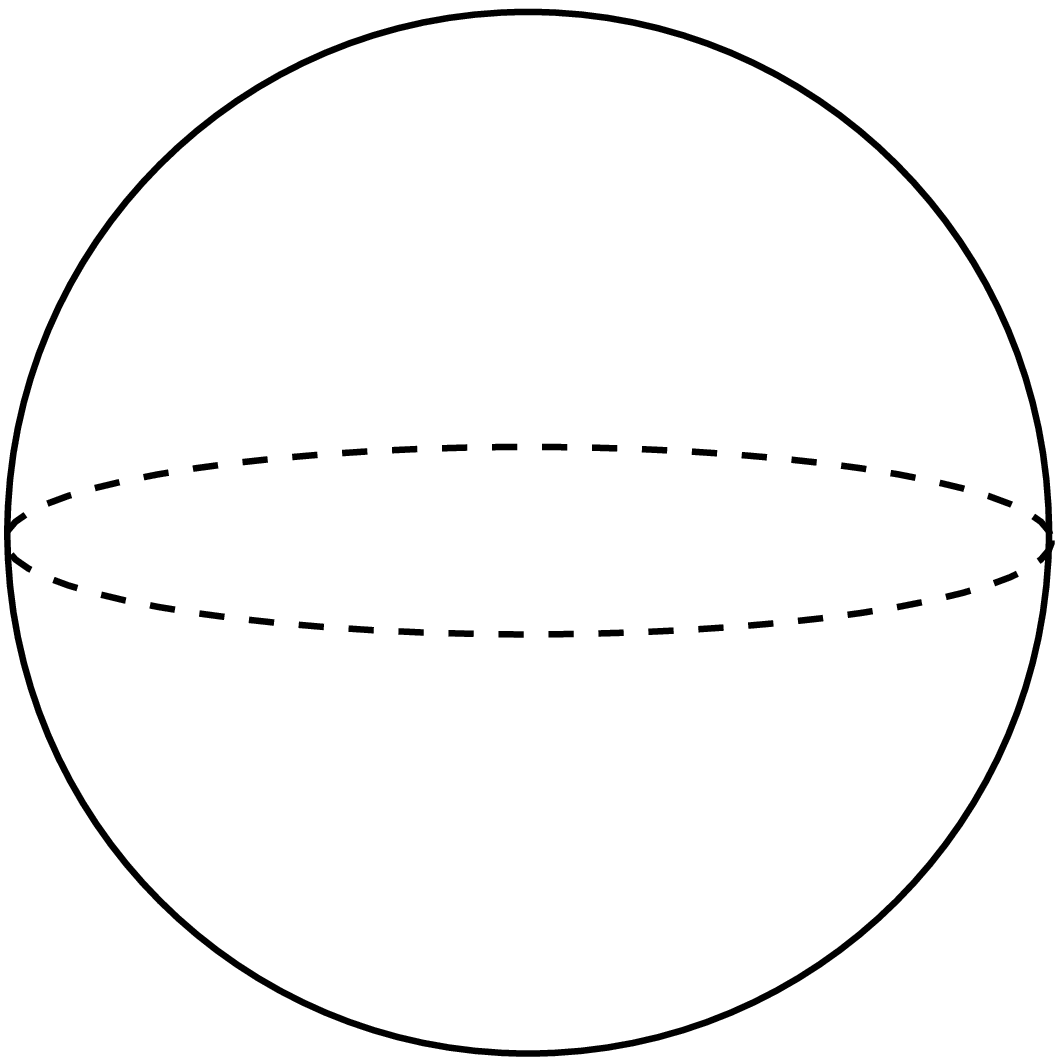}
\end{center}
\caption{Cartoon of the $AdS_5$ (bulk cylinder with boundary $R\times
  S^3$) space-time and the $S^5$ (sphere) space.}
\end{figure}

With the embedding coordinates $X^m(\tau,\sigma)$ and
$Y^m(\tau,\sigma)$ the Polyakov action of the \AdSS 
string in conformal gauge ($\eta_{ab}=\mbox{diag}(-1,1)$) takes the
form ($m,n=1,2,3,4,5$)
\be
\label{eq:AdSSaction}
I=-\frac{\sqrt{\lambda}}{4\pi}\, \int d\tau d\sigma\, \Bigl [ G_{mn}^{(AdS_5)}\, \partial_aX^m\, \partial^aX^n
+ G_{mn}^{(S^5)}\, \partial_a Y^m\, \partial^a  Y^n\, \Bigr ] + \mbox{fermions}\, ,
\ee
where we have suppressed the fermionic terms in the action, as they will not be relevant in our
discussion of classical solutions (the full fermionic action is stated in
\cite{hep-th/9805028,hep-th/0010104}). A natural choice of 
coordinates for the \AdSS space (``global coordinates'')
is
\footnote{This metric arises from a parametrization of the five sphere
$x_1^2+x_2^2+x_3^2+\ldots + x_6^2=1$ and anti-de-Sitter space
$-y_{-1}^2-y_0^2+y_1^2+y_2^2+\ldots y_4^2=-1$ through
\begin{align}
x_1+i\, x_2&= \sin\gamma\, \cos\psi \, e^{i\,\phi_1},\quad
x_3+i\, x_4= \sin\gamma\, \sin\psi \, e^{i\,\phi_2},\quad
x_5+i\, x_6= \cos\gamma\, \, e^{i\,\phi_3};\nn\\
y_1+i\, y_2&= \sinh\rho\, \cos\bar\psi \, e^{i\,\varphi_1},\quad
y_3+i\, y_4= \sinh\rho\, \sin\bar\psi \, e^{i\,\varphi_2},\quad
y_{-1}+i\, y_0= \cosh\rho\, \, e^{i\,t}\, .\nn
\end{align}
The embedding coordinates $X^m(\tau,\sigma)$ and $Y^m(\tau,\sigma)$ in
\eqn{eq:AdSSaction} are hence given
by $X^m=(\rho,\bar\psi,\varphi_1,\varphi_2,t)$ and $Y^m=(\gamma,\psi,\phi_1,\phi_2,\phi_3)$.}
\begin{align}
\label{eq:metrics}
ds^2_{AdS_5} &= d\rho^2 -\cosh^2 \rho \, dt^2 +\sinh^2\rho\, (\,
d\bar\psi^2 +\cos^2\bar\psi\, d{\varphi_1}^2
+\sin^2\bar\psi\, d{\varphi_2}^2\, )\nn\\
ds^2_{S^5} &= d\gamma^2 +\cos^2\gamma\, d{\phi_3}^2 +\sin^2\gamma\, (\, d\psi^2 +\cos^2\psi\, d{\phi_1}^2
+\sin^2\psi\, d{\phi_2}^2\, )\, .
\end{align}
Moreover, we have directly written the string
action with the help of the effective string tension $\sqrt{\lambda}=R^2/\alpha'$ 
of \eqn{eq:AdS/CFT rels}. It is helpful to picture the $AdS_5$
space-time as a bulk cylinder with a four dimensional boundary of the
form $\Real\times S^3$ (see figure \ref{fig:one}).
A consequence of the conformal gauge choice are the Virasoro
constraints, which take the form
\be
\label{VirConstr}
0\stackrel{!}{=} \dot X^m\, X'_m + \dot Y^p\, {Y'_p} \qquad
0\stackrel{!}{=} \dot X^m\, \dot X_m + \dot Y^p\, \dot Y_p 
+ {X^m}'\, X_m' +  {Y^p}'\,  Y_p' \, ,
\ee
where the dot refers to $\partial_\tau$ and the prime to
$\partial_\sigma$ derivatives. Of course the contractions in the
above are to be performed with the metrics of \eqn{eq:metrics}.

As mentioned in the introduction it is
presently unknown how to perform an exact quantization of this model.
It is, however, possible to perform a quantum fluctuation expansion
in $1/\sqrt{\lambda}$. For this one expands 
around a classical solution of \eqn{eq:AdSSaction} and integrates out
the fluctuations in the path-integral loop order by order. 
This is the route we will follow. Of course in doing so one will
only have a patch-wise access to the full spectrum of the theory,
with each patch given by the solution expanded around.

\subsection{The rotating point-particle}
\label{sect:particle}
It is instructive to sketch this procedure by considering the perhaps
simplest solution to the equations of motion of \eqn{eq:AdSSaction}:
the rotating point-particle on $S^5$, which is a degenerated string
configuration.

\begin{minipage}{6cm}
\begin{align}
\label{eq:pp}
t&=\kappa\,\tau \quad \rho=0 \quad \gamma=\frac{\pi}{2}\nn\\
\phi_1&=\kappa\,\tau \quad\phi_2=\phi_3=\psi=0\nn
\end{align}
\end{minipage}
\qquad\qquad
\begin{minipage}{6cm}
\raisebox{-2cm}{\epsfxsize=1.2cm\epsfbox{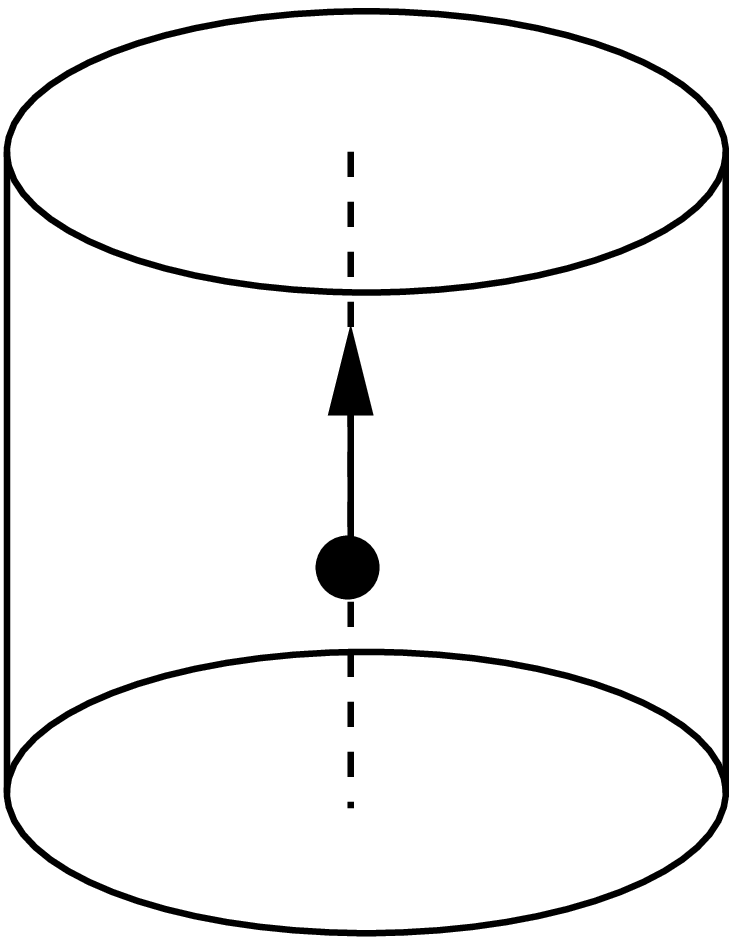}}
\raisebox{-1.3cm}{$\times$}
\raisebox{-2cm}{\epsfxsize=1.4cm\epsfbox{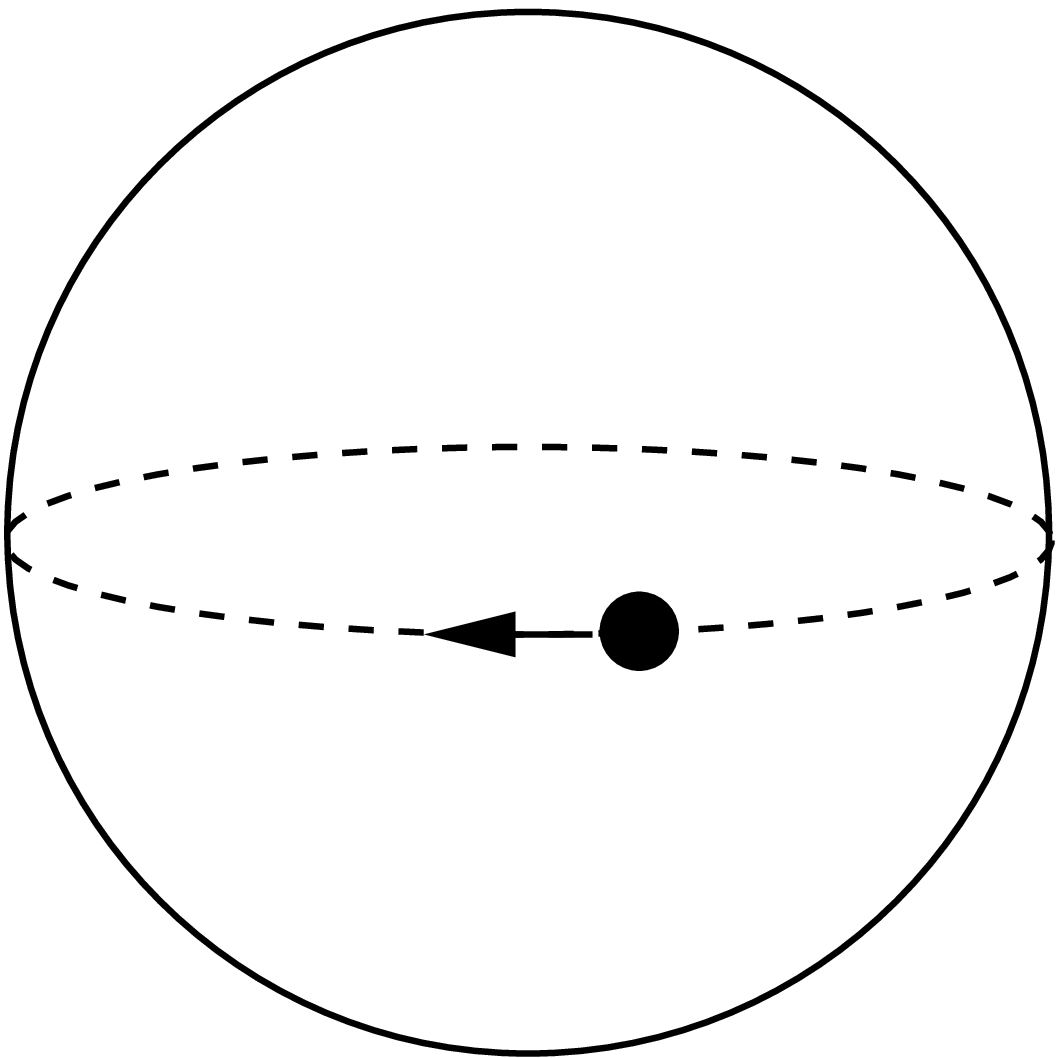}}\\[1cm]
\end{minipage}

One easily shows that this configurations satisfies the equations of
motion and the Virasoro constraints. 

A glance at
eqs.~\eqn{eq:AdSSaction} and \eqn{eq:metrics} reveals that the cyclic
coordinates of the action $I=\int dt\, L$ are
$(t,\varphi_1,\varphi_2;\phi_1,\phi_2,\phi_3)$
leading to the conserved charges $(E,S_1,S_2;J_1,J_2,J_3)$,
corresponding to the energy $E$ and two spins $(S_1,S_2)$ on 
$AdS_5$ as well as the three angular momenta $(J_1,J_2,J_3)$
on the five sphere respectively. The energy $E$ and
the first angular momentum $J_1$ are the only non vanishing conserved
quantities of the above point-particle configuration
and take the values (we also spell out $J_2$ for later use)
\bea
E&:=&\frac{\partial L}{\partial \dot t}=
\sqrt{\lambda}\int_0^{2\pi}\ft{d\sigma}{2\pi}\, 
\cosh^2\rho\, \dot t = {\sqrt{\lambda}}\, \kappa \, ,\nn\\
J_1&:=& -\frac{\partial L}{\partial \dot \phi_1}=
\sqrt{\lambda}\int_0^{2\pi}\ft{d\sigma}{2\pi}\,\sin^2\gamma\,
\cos^2\psi\,\dot \phi_1 = {\sqrt{\lambda}}\, \kappa\, , \label{EJ1J2}\\
J_2&:=& -\frac{\partial L}{\partial \dot \phi_2}=
\sqrt{\lambda}\int_0^{2\pi}\ft{d\sigma}{2\pi}\,\sin^2\gamma\,
\sin^2\psi\,\dot \phi_2 = 0\, ,\qquad
J:= J_1+J_2\, . \nn
\eea
Hence classically $E=J$. One may now consider quantum fluctuations
around this solution, i.e. $X^\mu=X^\mu_{\rm{solution}}(\tau) +
\frac{1}{\lambda^{1/4}}\, x^\mu(\tau,\sigma)$, and can integrate out
the quantum field $x^\mu(\tau,\sigma)$ in a perturbative fashion. 
This will result in `quantum' corrections to the classical energy $E_0$  in terms of
an  expansion in $1/\sqrt{\lambda}$
\be
E=\sqrt{\lambda}\, \kappa + E_2(\kappa) + \frac{1}{\sqrt{\lambda}}\,
E_4(\kappa) +\ldots
\ee
They key idea of Berenstein, Maldacena and Nastase \cite{hep-th/0202021} was to
consider the limit $J\to\infty$ with the parameter $\kappa=J/\sqrt{\lambda}$ held
fixed. This limit of a large quantum number 
suppresses all the higher loop contributions beyond one-loop, i.e.
\be
E-J= E_2(\kappa) + \frac{1}{\sqrt{J}}\, \tilde E_4(\kappa) +\ldots
\stackrel{J\to\infty}{\longrightarrow}  E_2(\kappa)\, .
\ee 
Hence the quadratic approximation becomes exact! This quadratic
fluctuation action (including the fermions) is nothing but the IIB
superstring in a plane wave background \cite{hep-th/0112044}, which
arises from the \AdSS geometry through a so-called Penrose limit
(see \cite{hep-th/0110242,hep-th/0201081} for this construction). The quantization of
this string model is straightforward in the light-cone gauge 
\cite{hep-th/0112044,hep-th/0202109} and leads to a free, massive two
dimensional theory for the transverse degrees of freedom ($i=1,\ldots,
8$)
\begin{equation}
I_2=\int d\tau d\sigma (\frac 12 \partial_ax^i\, \partial_ax^i
-\frac{\kappa^2}{2}\, x^i\, x^i + \mbox{fermions})
\end{equation}
with a compact expression for the spectrum 
\be
E_2=\frac{1}{\sqrt{\lambda'}}\,
\sum_{n=-\infty}^\infty\sqrt{1+\lambda'\, n^2}\, \widehat N_n
\qquad \lambda':= \frac{1}{\sqrt{\kappa}}=\frac{\lambda}{J^2}
\ee
where $\widehat N_n:=\alpha^{\dagger\, i}_n\, \alpha_n^i$ is the
excitation number operator for transverse  string excitations
$\alpha^{\dagger\, i}_n\, |0\rangle$ with
$[\alpha_m^i,\alpha^{\dagger\, j}_n]=\delta_{nm}\, \delta_{ij}$. The
Virasoro constraints \eqn{VirConstr} reduce to the level matching condition
$\sum_n n \, \widehat N_n=0$ known from string theory in flat Minkowski space-time. 
Hence the first stringy excitation is
$\alpha_n^\dagger\, \alpha_{-n}^\dagger\, |0\rangle$ with
$\sqrt{\lambda'}\, E_2=2\,
\sqrt{1+\lambda'\, n^2}$. For a more detailed treatment of the plane
wave superstring see \cite{hep-th/0307027,hep-th/0307101,hep-th/0310119,hep-th/0401155}.

\subsection{\Nfour Super Yang-Mills}

The conjectured dual gauge theory of the $AdS_5\times S^5$ superstring is
the maximally supersymmetric (${\cal N}=4$)
Yang-Mills theory in four dimensions \cite{NUPHA.B122.253,NUPHA.B121.77}.
Its field content is given by a gluon field $A_\mu(x)$, six scalars $\phi_i(x)$ 
($i=1,\ldots,6$) as well as 4 Majorana gluinos, which we choose to write as a 16 component
10d Majorana-Weyl spinor  $\chi_\alpha(x)$ ($\alpha=1,\ldots 16$). 
All fields are in the adjoint representation of $SU(N)$.
The action of ${\cal N}=4$ Super Yang-Mills is uniquely determined by two parameters, the
coupling constant $g_{\rm YM}$ 
and the rank of the gauge group $SU(N)$
\be
S=\frac{2}{\gym^2}\, \int d^4x\, \Tr\, \Bigl\{ \frac{1}{4}\, (F_{\mu\nu})^2
+\frac{1}{2}\, (D_\mu\phi_i)^2 -\frac{1}{4}\, [\phi_i,\phi_j]\,
[\phi_i,\phi_j] + \frac{1}{2}\, \bar\chi D\!\!\!\!/\,\, \chi - \frac{i}{2}
\, \bar \chi\, \Gamma_i\, [\phi_i,\chi]\, \Bigr \}
\ee
with the covariant derivative $D_\mu= \partial_\mu
-i[A_\mu,\,\,\,]$. Furthermore, $(\Gamma_\mu,\Gamma_i)$ are the
ten dimensional Dirac matrices.

Due to the large amount of supersymmetry present, the conformal invariance 
of the classical field theory survives the quantization procedure: The
coupling constant $\gym$ is not renormalized and its $\beta$-function
vanishes to all orders in perturbation theory
\cite{PHLTA.B100.245,NUPHA.B236.125,PHLTA.B123.323}.
This is why one often refers to ${\cal N}=4$ Super Yang-Mills
as a ``finite'' quantum field theory. Nevertheless composite gauge invariant
operators, i.e.~traces of products of fundamental 
fields and their covariant derivatives at the same space-point, 
e.g.~${\cal O}_{i_1\ldots i_k}(x)=\Tr[\phi_{i_1}(x)\,\phi_{i_2}(x)\ldots 
\phi_{i_k}(x)]$, are renormalized and acquire anomalous dimensions. These
may be read off from the two point functions (stated here for the case
of scalar operators)
\be
\langle {\cal O}_A(x)\, {\cal O}_B(y)\rangle = \frac{\delta_{AB}}{(x-y)^{2\,
\Delta_{{\cal O}_A}}}
\label{CFT2PtFct}
\ee
where $\Delta_{{\cal O}_A}$ is the  scaling dimension of the composite
operator ${\cal O}_A$. Classically these scaling dimensions are simply
the sum of the individual dimensions 
of the constituent fields ($[\phi_i]=[A_\mu]=1$ and $[\chi]=3/2$). 
In quantum theory the scaling dimensions 
receive anomalous corrections, organized in
a double expansion in $\lambda=\gym^2\, N$ (loops) and $1/N^2$ (genera)
\be
\Delta = \Delta_0 + \sum_{l=1}^\infty \lambda^l\, \sum_{g=0}^\infty
\frac{1}{N^{2\, g}}\, \Delta_{l,g} \, .
\label{anexp}
\ee
Determining the scaling dimensions in perturbation theory is a difficult task
due to the phenomenon of operator mixing: One has to identify the correct
basis of (classically) degenerate gauge theory operators $\cO_A$ in which
\eqn{CFT2PtFct} indeed becomes diagonal. This task is greatly facilitated
through the use of the gauge theory dilatation operator, to be discussed in
section \ref{sec:gt}.

The core statement of the AdS/CFT correspondence is that the scaling dimensions
$\Delta(\lambda,1/N^2)$ are equal to the energies $E$ of the \AdSS string excitations.
A central problem, next to actually computing these quantities on either side of the 
correspondence, is to establish a 'dictionary' between states in the string
theory and their dual gauge theory operators. Here the underlying symmetry structure
of $SU(2,2|4)$ is of help, whose bosonic factors are $SO(2,4)\times SO(6)$. 
$SO(2,4)$ corresponds to the isometry group of $AdS_5$  or 
the conformal group  in four dimensions respectively. $SO(6)$, on the
other hand, emerges from the isometries of the five sphere and the 
$R$-symmetry group of internal rotations of the six scalars and four gluinos in 
SYM. Clearly then any state or operator can be labeled by the eigenvalues of
the six Cartan generators of $SO(2,4)\times SO(6)$
\be
(E;\underbrace{S_1,S_2}_{S^3}; \underbrace{J_1,J_2,J_3}_{S^5})
\ee
where we shall denote the $S_i$ as the commuting 'spins' on the three sphere within 
$AdS_5$, the $J_k$ as the commuting angular momenta on the $S^5$ and $E$ is the 
total energy. These are the conserved quantities of the string
discussed above eq.~\eqn{EJ1J2}.

The strategy is now to search for string solutions with
energies 
$$E=E(S_1,S_2,J_1,J_2,J_3)$$ 
and to identify these with scaling dimensions of Super Yang-Mills operators carrying the
identical set of charges ($S_1$,$S_2$,$J_1$,$J_2$,$J_3$). Generically the string energies
will depend on additional data, such as winding numbers or oscillator levels, which we have
suppressed in the above.


\section{Spinning string solutions}
\label{section:spinning_string_solutions}

\begin{figure}[t]
\begin{center}
\includegraphics[width=4cm]{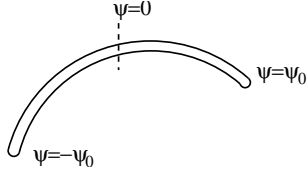}
\end{center}
\caption{The folded sting extending from $\psi=-\psi_0$ to
  $\psi=\psi_0$, where $\sin^2\psi_0:=q$.}
\label{fig:foldedstring}
\end{figure}

We shall now look for a solution of the \AdSS string with $S_1=S_2=0$ and $J_3=0$,
i.e.~a string configuration rotating in the $S^3$ within $S^5$
and evolving only with the time coordinate of the $AdS_5$ space-time. This was first
discussed by Frolov and Tseytlin in \cite{hep-th/0304255,hep-th/0306143}. 
For this let us consider the
following ansatz in the global coordinates of
\eqn{eq:metrics} 
\be
t=\kappa\, \tau \qquad \rho=0 \qquad \gamma= \ft \pi 2 \qquad
\phi_3=0  \qquad \phi_1=\omega_1\, \tau \qquad  \phi_2=\omega_2\, \tau
\qquad \psi=\psi(\sigma)\, ,
\label{ansatz}
\ee 
with the constant parameters $\kappa$, $\omega_{1,2}$ and the profile $\psi(\sigma)$ to be
determined. The string action \eqn{eq:AdSSaction} then becomes 
\be
I=-\frac{\sqrt\lambda}{4\pi}\, \int d\tau\, \int_0^{2\pi}d\sigma\, 
\Bigl [ \, \kappa^2 + {\psi'}^2 - \cos^2\psi\, {\omega_1}^2 - \sin^2\psi\, {\omega_2}^2 \,
\Bigr ]\, ,
\ee
leading to an equation of motion for $\psi(\sigma)$
\be
\psi'' + \sin\psi\, \cos\psi\, ({\omega_2}^2-{\omega_1}^2) =0\, .
\label{EOMpsi}
\ee
We define ${\omega_{21}}^2:= {\omega_2}^2-{\omega_1}^2$, which we take to be positive
without loss of generality. Integrating this equation once yields the ``string pendulum''
equation
\be
\frac{d\psi}{d\sigma} = \omega_{21}\,\sqrt{q - \sin^2\psi}\, ,
\label{stringpendulum}
\ee
where we have introduced an integration constant $q$. Clearly there are two qualitatively
distinct situations for $q$ larger or smaller than one: For $q\leq 1$ we have a folded string
with $\psi$ ranging from $-\psi_0$ to $\psi_0$, where $q=\sin^2\psi_0$, and $\psi'=0$ at
the turning points where the string folds back onto itself (see figure
\ref{fig:foldedstring}).
If, however, $q>1$ then $\psi'$ never vanishes and we have a circular
string configuration embracing a full circle on the $S^3$: The energy stored in the system is
large enough to let the pendulum overturn.

In addition we have to fulfill the Virasoro constraint equations \eqn{VirConstr}. One checks that
our ansatz \eqn{ansatz} satisfies the first constraint of \eqn{VirConstr}, whereas the second
constraint equation leads to
\be
q=\frac{\kappa^2 -{\omega_1}^2}{{\omega_{21}}^2}\qquad \qquad({\omega_{21}}^2\neq 0)\, ,
\label{F1}
\ee
relating the integration constant $q$ to
the parameters of our ansatz\footnote{The case $\omega_{21}=0$
is discussed in subsection \ref{sect:simplesol} below.}.
Our goal is to 
compute the energy $E$ of these two solutions as a function of the commuting angular momenta
$J_1$ and $J_2$ on the three sphere within $S^5$. Upon using
eqs.~\eqn{EJ1J2} and inserting the ansatz \eqn{ansatz} these are given by
\bea
E&=& \sqrt{\lambda} \, ,\\
\label{J1J2def}
J_1&=&  \sqrt{\lambda}\,\omega_1\,\int_0^{2\pi}
\ft{d\sigma}{2\pi}\,\cos^2\psi(\sigma)\, ,\qquad
J_2=  \sqrt{\lambda}\, \omega_2\,\int_0^{2\pi}
\ft{d\sigma}{2\pi}\,\sin^2\psi(\sigma)  \, .
\eea
From this we learn that
\be
\label{F3}
\sqrt{\lambda}= \frac{J_1}{\omega_1}+ \frac{J_2}{\omega_2} \, .
\ee
\subsection{The special case: $\omega_1=\omega_2$}
\label{sect:simplesol}
It is instructive to first discuss the particularly simple special case of a circular 
string, where $\omega_1=\omega_2$ and \eqn{F1} does not apply. This
will turn out to be a limiting case of the $q>1$ scenario. For $\omega_{21}=0$
 the equation of motion for $\psi(\sigma)$ \eqn{EOMpsi} immediately yields 
\be
\psi''=0 \qquad \Rightarrow \qquad \psi(\sigma)=n\, \sigma
\ee
with
$n$ being the integer winding number of this ciruclar string $\psi(\sigma+2\pi)=
\psi(\sigma)+2\pi\, n$.
In this case the Virasoro constraints \eqn{VirConstr} yield 
$\kappa=\sqrt{n^2+{\omega_1}^2}$.
A little bit of algebra quickly shows that the energy $E$ may be reexpressed as
a function of $J_1=J_2$ and reads
\be
\label{C1en}
E=2\, J_1\, \sqrt{1+\frac{n^2\, \lambda}{4\, {J_1}^2}}
\ee
which is analytic in $\frac{\lambda}{{J_1}^2}$! This amounts to an all loop prediction for
the dual gauge theory scaling dimension in the BMN limit $J_1\to\infty$ with $\lambda/{J_1}^2$
fixed, quite similar to the result for the plane-wave superstring discussed above.

Let us now discuss the folded and circular string solutions in turn.

\subsection{The folded string: $q\leq 1$}
\label{subsect:3.2}

In the folded case $J_1$ may be expressed in terms of an elliptic
integral\footnote{Our conventions are ($x<1$) 
$$
E(x):=\int_0^{\pi/2}d\psi \, \sqrt{1-x\, \sin^2\psi} \qquad
K(x):=\int_0^{\pi/2}d\psi \, \frac{1}{\sqrt{1-x\, \sin^2\psi}}\, .
$$} by substituting (using \eqn{stringpendulum})
\begin{equation}
d\sigma=\frac{d\psi}{\omega_{21}\sqrt{q-\sin^2\psi}}
\end{equation}
into \eqn{J1J2def}
and performing some elementary transformations to find ($q=\sin^2\psi_0$) 
\be
\label{F2}
J_1=\frac{\sqrt{\lambda}\, \omega_1}{2\pi}\, 4\, \int_0^{\psi_0}
  d\psi\, \frac{\cos^2\psi}{\omega_{21}\, \sqrt{\sin^2\psi_0 -\sin^2\psi}}
= \frac{2\, \sqrt{\lambda}\, \omega_1}{\pi\, \omega_{21}}\,
E(\sin^2\psi_0)\, ,
\ee
where we only need to integrate over one quarter of the folded string
due to symmetry considerations (see figure \ref{fig:foldedstring}). 
Additionally we have
\be
\label{F4}
2\pi= \int_0^{2\pi}d\sigma = 4\,
\int_0^{\psi_0}\frac{d\psi}{\omega_{21}\, \sqrt{\sin^2\psi_0
    -\sin^2\psi}} = \frac{4}{\omega_{21}}\, K(\sin^2\psi_0) \, .
\ee
The four equation \eqn{F1}, \eqn{F3}, \eqn{F2} and \eqn{F4} may then be
used to eliminate the parameters of our solution $\kappa$,
$\omega_1$ and $\omega_2$. For this rewrite \eqn{F2} and \eqn{F4} as ($\kappa=E/\sqrt{\lambda}$)
\be
\label{omegas}
\frac{\omega_1}{\omega_{21}}=\frac{\pi}{2\,\sqrt{\lambda}}\, \frac{J_1}{E(q)}\, ,\qquad
\frac{\kappa}{\omega_{21}} = \frac{\pi}{2\, \sqrt{\lambda}}\, \frac{E}{K(q)} \, ,
\ee
and use \eqn{F3} to deduce
\be
\label{omegas2}
\frac{\omega_2}{\omega_{21}}= \frac{\pi}{2\, \sqrt{\lambda}}\, \frac{J_2}{K(q)-E(q)} \, .
\ee
Then the Virasoro constraint equation \eqn{F1} and the identity $1=({\omega_2}^2-{\omega_1}^2)/{\omega_{21}}^2$
yield the two folded string equations
\be
\frac{4\, q\, \lambda }{\pi^2}= \frac{E^2}{K(q)^2} -
\frac{{J_1}^2}{E(q)^2}\, , \qquad
\frac{4\, \lambda}{\pi^2} = \frac{{J_2}^2}{(K(q)-E(q))^2} -
\frac{{J_1}^2}{E(q)^2}\, ,
\label{Fstringeqs}
\ee
which implicitly define the sought after energy function $E=E(J_1,J_2)$ upon further elimination of
$q$. This is achieved by assuming an analytic behavior of
$q$ and $E$ in the BMN type limit of large total angular 
momentum $J:=J_1+J_2\to\infty$ with $\lambda/J^2$ held fixed
\begin{align}
\label{BMNexp}
q&=q_0+\frac{\lambda}{J^2}\, q_1 + \frac{\lambda^2}{J^4}\, q_2 +\ldots
\nn\\
E&= J\, \Bigl (E_0 + \frac{\lambda}{J^2}\, E_1 +
\frac{\lambda^2}{J^4}\, E_2 +\ldots \Bigr ) \, .
\end{align}
Plugging these expansions into \eqn{Fstringeqs} one can solve for the $q_i$ and
$E_i$ iteratively. At leading order $E_0=1$ (as it should from the
dual gauge theory perspective) and $q_0$ is implicitly determined through
the ``filling fraction'' $J_2/J$
\be
\label{FE1}
 \frac{J_2}{J} = 1- \frac{E(q_0)}{K(q_0)} \, .
\ee
The first non-trivial term in the energy is then  expressed in terms of
$q_0$ through
\be
\label{FEprediction}
E_1=\frac{2}{\pi^2}\, K(q_0)\, \Bigl ( \, E(q_0) - (1-q_0)\, K(q_0)\,
\Bigr )
\ee
yielding a clear prediciton for one-loop gauge theory. The higher
order $E_i$ can then also be straightforwardly
obtained. To give a
concrete example, let us evaluate $E$ for the first few orders in
$\lambda/J^2$ in the ``half filled'' case $J_1=J_2$:
\be
E=2\, J_1\, ( 1 + \frac{0.71}{8}\, \frac{\lambda}{{J_1}^2} -
\frac{1.69}{32}\,  \frac{\lambda^2}{{J_1}^4} + \ldots )
\ee
which shows that the energy of the folded string configuration is
smaller than the one of the closed configuration of \eqn{C1en} for
single winding $n=1$.

\subsection{The circular string: $q> 1$}
\label{subsect:3.3}

For $q>1$ the string does not fold back onto itself and extends around
the full circle of $\psi\in[0,2\pi]$
as $\sigma$ runs from $0$ to $2\pi$. 
The initial expression for \eqn{F2} now changes only marginally to
\be
J_1=\frac{\sqrt{\lambda}\, \omega_1}{2\pi}\, \int_0^{2\pi}
  d\psi\, \frac{\cos^2\psi}{\omega_{21}\, \sqrt{q -\sin^2\psi}} \, .
\ee
Elementary transformations yield
\be
\label{C2}
J_1= \frac{2\, \sqrt{\lambda}\, \omega_1}{\pi\, \omega_{21}}\Bigr [
\, \frac{1-q}{\sqrt{q}}\, K(q^{-1}) + \sqrt{q}\, E(q^{-1})\, \Bigr ] \, .
\ee
Analogously \eqn{F4} now becomes
\be
\label{C4}
2\pi = \int_0^{2\pi}d\sigma = 
4\,
\int_0^{\pi/2}\, \frac{d\psi}{\omega_{21}\, \sqrt{\sin^2\psi_0
    -\sin^2\psi}} = \frac{4}{\omega_{21}\,\sqrt{q}}\, K(q^{-1}) \, .
\ee
The corresponding relations to \eqn{omegas} and \eqn{omegas2} then take  the form
\begin{align}
\frac{1}{\omega_{21}} &= \frac{\pi}{2}\, \frac{\sqrt{q}}{K(q^{-1})} \, , \qquad
\frac{\omega_1}{\omega_{21}}=\frac{\pi}{2\,\sqrt{\lambda}}\, 
\frac{\sqrt{q}\, J_1}{(1-q)\, K(q^{-1})+
q\, E(q)}\, ,\nn\\
\frac{\omega_2}{\omega_{21}}&= \frac{\pi}{2\, \sqrt{\lambda}}\, 
\frac{J_2}{\sqrt{q}[\, K(q^{-1})-E(q^{-1})\, ]} \, .
\end{align}
From these one deduces in complete analogy to \eqn{Fstringeqs} the two
circular string equations
\begin{align}
\frac{4\, \lambda }{\pi^2}&= \frac{E^2}{K(q^{-1})^2} -
\frac{{J_1}^2}{[\, (1-q)\, K(q^{-1})+q\, E(q^{-1})\, ]^2}\nn\\
\frac{4\, q\, \lambda}{\pi^2} &= \frac{{J_2}^2}{[\,K(q^{-1})-E(q^{-1})\,]^2} -
\frac{q^2\,{J_1}^2}{[\,(1-q)\, K(q^{-1})+q\, E(q^{-1})\, ]^2} \, ,
\label{Cstringeqs}
\end{align}
which encode the energy relation $E=E(J_1,J_2)$ upon elimination of $q$.
In order to do this we again make an analytic ansatz in $\lambda/J^2$ for $q$ and $E$ 
as we did in \eqn{BMNexp}. This yields
the following implicit expression for $q_0$ in terms of the filling fraction $J_2/J$
\be
\label{CS1}
 \frac{J_2}{J} = q_0\, \Bigl ( 1- \frac{E(q_0^{-1})}{K(q_0^{-1})} \Bigr )\, ,
\ee
The first two energy terms in the $\lambda/J^2$ expansion then take the form
\be
\label{CS2}
E_0=1\, , \qquad E_1= \frac{2}{\pi^2}\, E(q_0^{-1})\, K(q_0^{-1}) \, ,
\ee
which again gives a clean prediction for the dual gauge theory scaling dimensions
at one-loop.

\begin{figure}[t]
\begin{center}
\includegraphics[width=9cm]{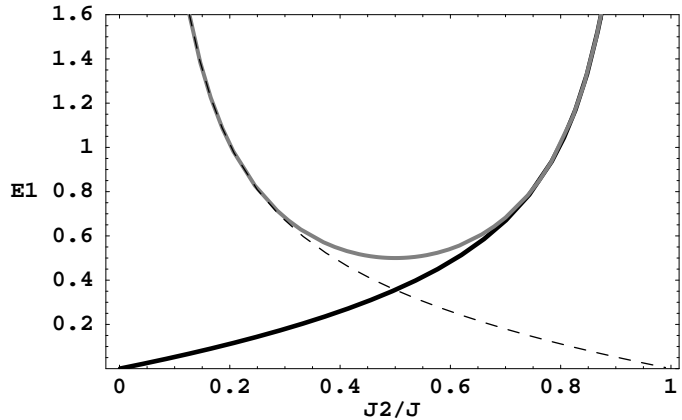}
\end{center}
\caption{The one-loop energies of the folded (dark) and circular (light) string solutions
plotted against the filling fraction $J_2/J$. The dashed curve is the mirrored folded string
solution where one interchanges $J_1\leftrightarrow J_2$.} 
\label{Eplot}
\end{figure}

In figure~\ref{Eplot} we have plotted the energies of the folded and circular string solutions
against the filling fraction $J_2/J$. As expected from the ``string pendulum'' picture of 
\eqn{stringpendulum} the folded string solution has lower energy for fixed filling
fraction $J_2/J$: The pendulum
does not perform a full turn, but oscillates back and forth. Note that the folded string
approaches $E_1=0$ in the limit $J_2/J\to 0$, which is the rotating point-particle solution
of subsection~\ref{sect:particle}.  The quantum fluctuations about it correspond to 
the plane-wave string domain. Note also that the simplest circular string solution with $\omega_1=\omega_2$
discussed in subsection~\ref{sect:simplesol} is the minimum ($J_2/J=0.5$) of the full circular
string solution for $n=1$. 

These classical string energies will be reproduced in a dual one-loop gauge theory computation
in section~\ref{sec:gt}.

\subsection{Further Developments}

The discussed explicit solutions of the bosonic string on a $\Real\times S^3$
background are the first simple elements of the general set of
classical solutions of the bosonic $AdS_5\times S^5$  string
considered in the literature. 
Earlier examples not discussed in the text are 
\cite{hep-th/0204051,hep-th/0205244,hep-th/0209047}.
The bosonic part of the classical string action consists of a $SO(6)$
and a $SO(2,4)$ $\sigma$-model augmented by the conformal gauge
(Virasoro) constraints. The $O(p,q)$ $\sigma$-models have been known 
to be integrable for a long time \cite{CMPHA.46.207,NUPHA.B137.46,Zakharov:1973pp,Faddeev:1985qu} and
one would expect this to remain true also once one imposes the
conformal gauge constraints. And indeed in
\cite{hep-th/0307191,hep-th/0311004} more complicated solutions of the
$AdS_5\times S^5$ string were constructed by reducing the system to
the so-called Neumann integrable system through a suitable ansatz.
The solutions of \cite{hep-th/0307191,hep-th/0311004}
involve nonvanishing values for all spins and
angular momenta $(S_1,S_2,J_1,J_2,J_3)$. However, they do not
generically display an analytic behavior in the effective coupling
constant $\lambda'$. This is true only for configurations with at
least one large charge $J_i$ on the $S^5$.

In the context of the $O(p,q)$ $\sigma$-models integrability is based
on the existence of a Lax pair, a family of flat connections on the 2d string
worldsheet, giving rise to an infinite number of conserved
charges. These were first discussed in the context of the bosonic string
in \cite{hep-th/0206103} and for the full superstring in \cite{hep-th/0305116}.
The Lax pair for the string was put to use in \cite{hep-th/0402207} for
string configurations on $\Real\times S^3$ -- the sector we also considered in
the above. These investigations led to the construction of an
underlying algebraic curve parametrizing the solutions.
This enabled the authors of \cite{hep-th/0402207} to write down an integral equation of Bethe type 
yielding the associated energies of the solutions. Very similar equations
will appear below in our discussion in section~\ref{subsect:thermo}
on the thermodynamic limit of the gauge theory Bethe equations. 
The extraction of these integral equations from the string
$\sigma$-model then allows for a direct comparison to the gauge
theory Bethe equations. On this level of formalization, there is no
need to compare explicit solutions any longer -- as we are doing here for
pedagogical purposes. This construction based on an underlying algebraic
curve makes full use of the technology of integrable systems and has been nicely 
reviewed by Zarembo in \cite{hep-th/0411191}.

In the very interesting paper \cite{hep-th/0406256} these continuum string
Bethe equations were boldly discretized  leading to a conjectured set of
Bethe equations for the {\sl quantum} spectrum of the string. 
This proposal has been verified by comparing it to known quantum data 
of the $AdS_5\times S^5$ string: 
The near plane-wave spectrum of the superstring of 
\cite{hep-th/0307032,hep-th/0404007,hep-th/0405153,hep-th/0407096}
as well as the expected \cite{hep-th/9802109} generic scaling of the string energies with
$\lambda^{1/4}$ in the strong coupling limit agree with the
predictions of the quantum string Bethe equations. But there is more
quantum data for the $AdS_5\times S^5$ string available: In a series of papers by
Tseytlin, Frolov and collaborators one-loop corrections on the string
worldsheet to the energies of various spinning string solutions have been computed 
\cite{hep-th/0306130,hep-th/0408187,hep-th/0501203}. The one loop
correction for a circular string moving in $AdS_3\times S^1 \subset
AdS_5\times S^5$ obtained in \cite{hep-th/0501203} was recently
compared \cite{hep-th/0507189} to the
result obtained from the proposed quantum string Bethe equations of
\cite{hep-th/0406256}. 
The authors of \cite{hep-th/0507189} find agreement 
when they expand the results in $\lambda'$ (up to third order), but
disagreements emerge in different
limits (where $\lambda'$ is not small). The interpretation of this
result is unclear at present.
Finally the proposed
quantum string Bethe equations of \cite{hep-th/0406256} can also 
be microscopically attributed
to a $s=1/2$ spin chain model with long-range interactions up
 to (at least) order five in a small $\lambda$ expansion \cite{hep-th/0409054}.

The technically involved construction of algebraic curves solving the
classical $\Real\times S^3$ string $\sigma$-model 
has subsequently been generalized to larger sectors: In
\cite{hep-th/0410253} to $\Real\times S^5$ (or $SO(6)$ in gauge theory language) configurations,
in \cite{hep-th/0410105} to $AdS_3\times S^1$ (or $SL(2)$) string configurations
and finally in \cite{hep-th/0502226} to superstrings propagating in
the full $AdS_5\times S^5$ space.

There has also been progress on a number of possible paths towards the
true quantization of the classical integrable model of the \AdSS string in the works
\cite{hep-th/0410282,hep-th/0411089,hep-th/0502240,hep-th/0504144,hep-th/0409159,hep-th/0411170}, 
however, it is fair to say that this problem remains currently unsolved. 

\section{The dual gauge theory side}
\label{sec:gt}

Let us now turn to the identification of the folded
and circular string solutions in the dual gauge theory.

Our aim is to reproduce the obtained energy functions $E_1(J_1,J_2)$ 
 plotted in figure \ref{Eplot} 
from a dual gauge theory computation
at one-loop. For this we need to identify the gauge theory operators,
which are dual to the spinning strings on $\Real\times S^3$.
As here $J_2=0=S_1=S_2$ the relevant operators will
be built from the two complex scalars
$Z:=\phi_1+i\,\phi_2$ and $W:=\phi_3+i\,\phi_4$ with a total number of
$J_1$ $Z$-fields and $J_2$ $W$-fields, i.e.
\be
{\cal O}^{J_1,J_2}_\alpha= \Tr[\, Z^{J_1}\, W^{J_2}\, ] +\ldots \, ,
\label{calO} 
\ee
where the dots denote suitable permutations of the $Z$ and $W$ to be discussed. 
An operator of the form \eqn{calO}
may be pictured as a ring of black (``$Z$'') and red (``$W$") beads -- or equivalently as a
configuration of an $s=1/2$ quantum spin chain, where $W$ corresponds
to the state
$|\uparrow\,\rangle$ and $Z$ to $|\downarrow\,\rangle$. 
$$
\Tr[Z W^2 Z W^4 ] \quad \Leftrightarrow \quad
\raisebox{0.5cm}{\rotatebox{-90}{{\epsfysize=1.4cm\epsfbox{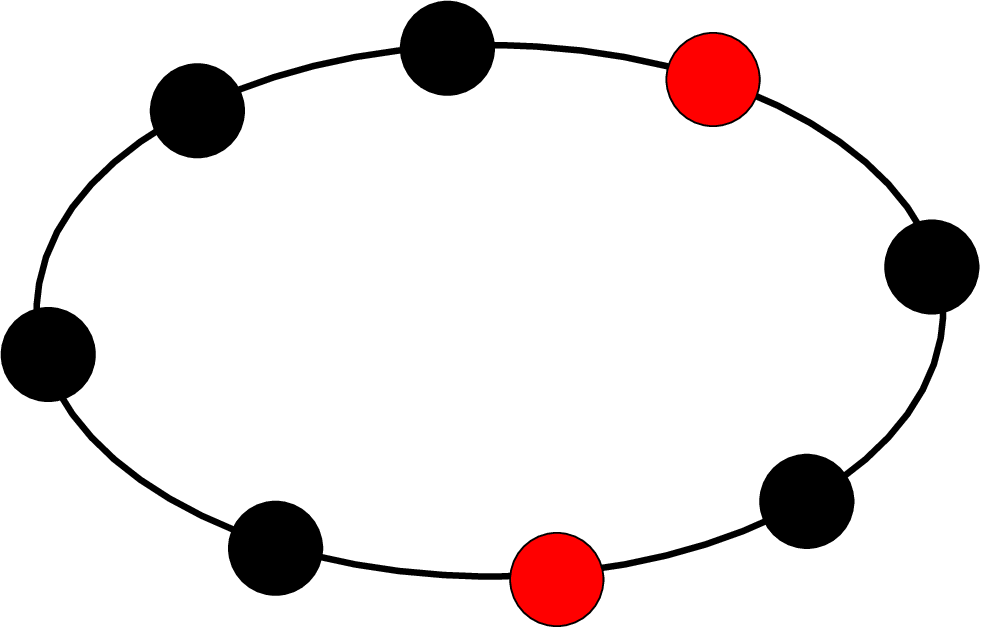}}}}
\quad \Leftrightarrow \quad 
|\downarrow\,\uparrow\,\uparrow\,\downarrow\,\uparrow\,\uparrow
\,\uparrow\,\uparrow\,\rangle_{\rm cyclic}
$$

How does one compute the associated scaling dimensions at (say) one
loop order for $J_1,J_2\to\infty$? Clearly one is facing a huge operator mixing
problem as all  ${\cal O}^{J_1,J_2}_i$ with arbitrary permutations of $Z$'s and $W$'s
 are degenerate at tree level where $\Delta^{{\cal O}^{J_1,J_2}_i}_0= J_1+J_2$. 

A very efficient tool to deal with this problem is 
the dilatation operator ${\cal D}$, which was introduced in 
\cite{hep-th/0212269,hep-th/0303060}. It acts on the trace
operators ${\cal O}^{J_1,J_2}_\alpha$ at a fixed space-time point $x$ and its eigenvalues are
the scaling dimensions $\Delta$
\be
{\cal D}\circ {\cal O}^{J_1,J_2}_\alpha(x) = \sum_{\beta}\, {\cal
  D}_{\alpha \beta}\, {\cal O}^{J_1,J_2}_\beta(x) \, .
\ee 
The dilatation operator is constructed in such a fashion as to attach the relevant 
diagrams to the open legs of the
\begin{figure}[t]
\begin{center}
$\displaystyle
{\cal D}\circ \raisebox{-.5cm}{\epsfysize=1.2cm\epsfbox{flatcirc0.eps}}
=
\raisebox{-.5cm}{\epsfysize=1.2cm\epsfbox{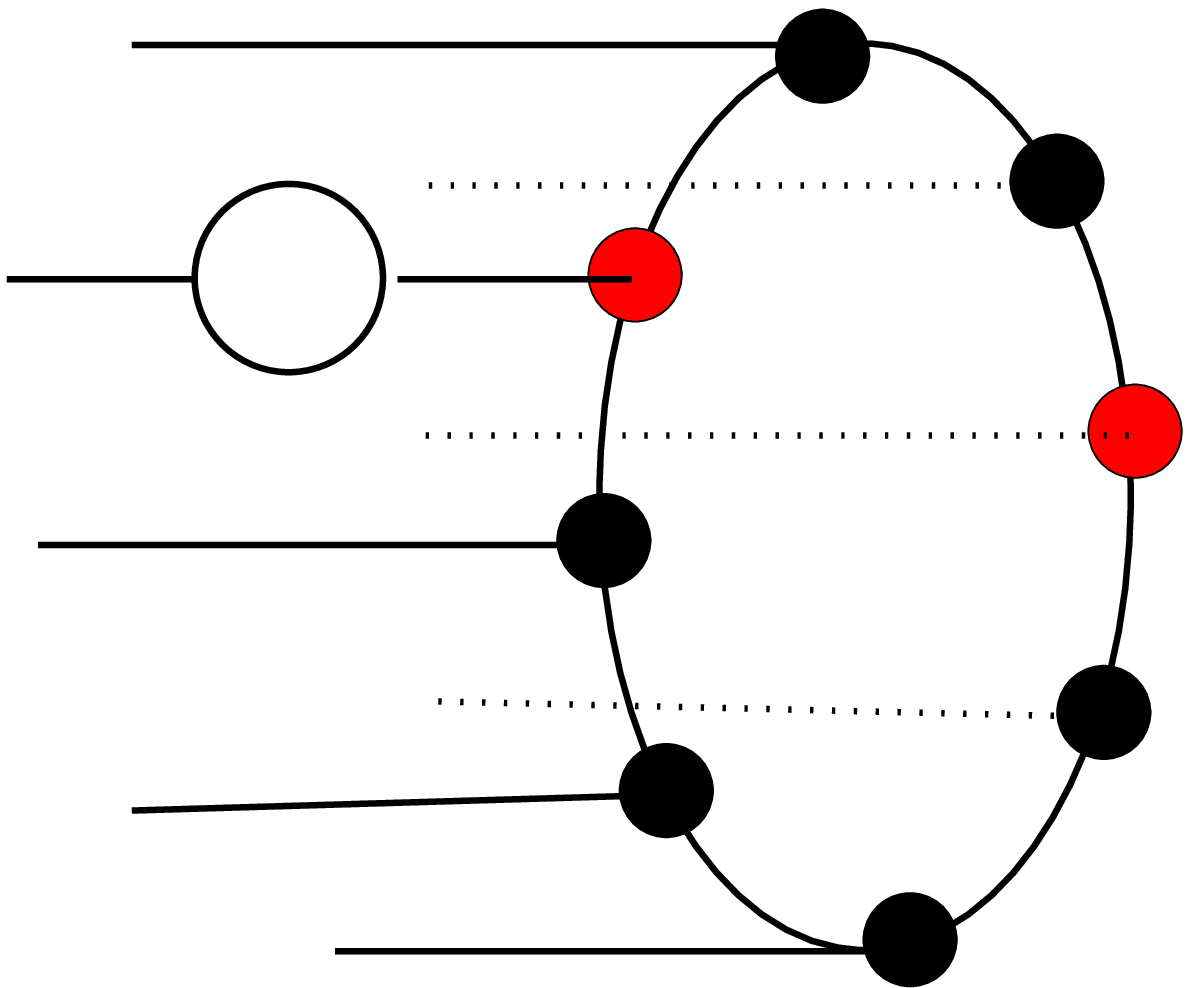}}\, +\,
\raisebox{-.5cm}{\epsfysize=1.2cm\epsfbox{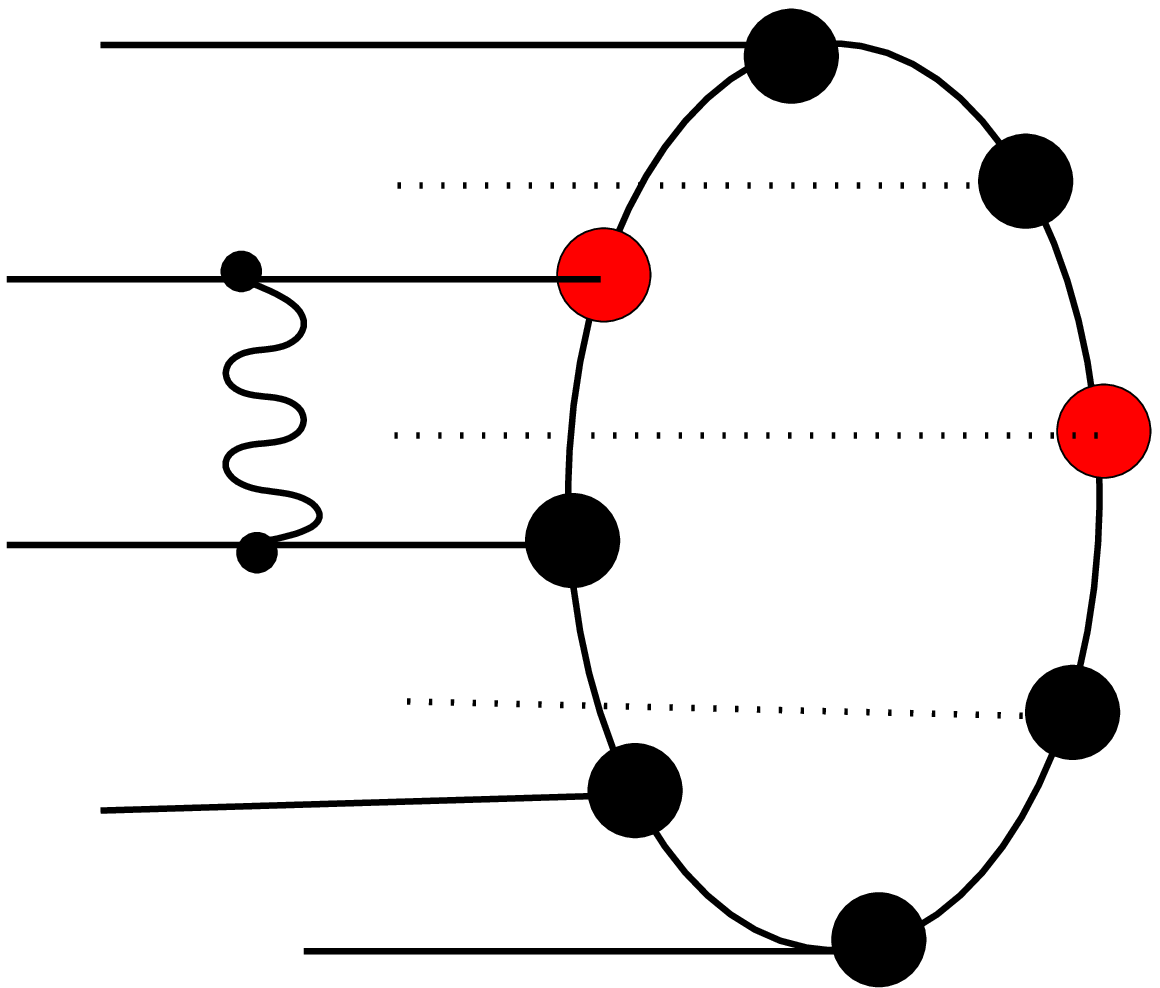}}\, +\,
\raisebox{-.5cm}{\epsfysize=1.2cm\epsfbox{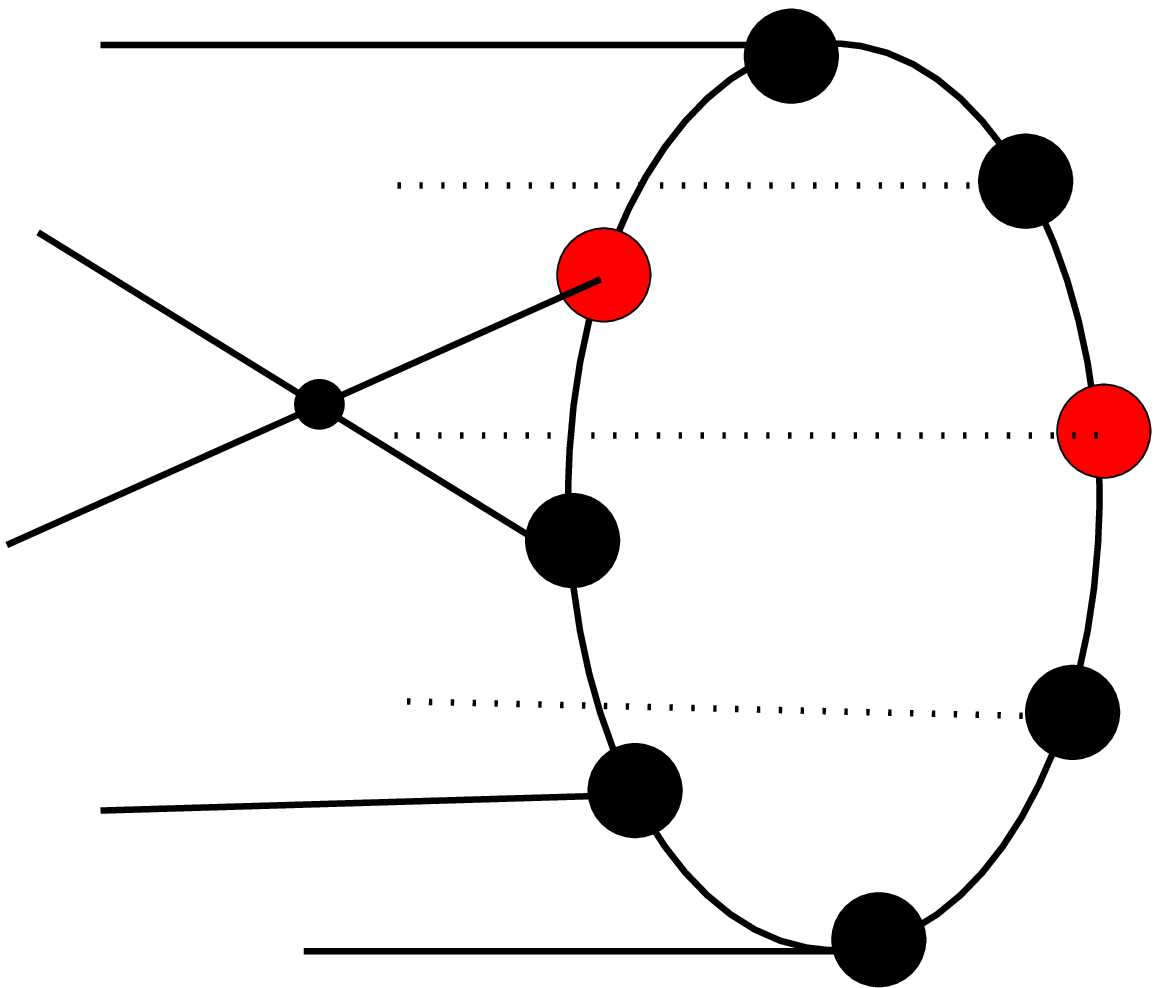}}\,+\ldots
=
\sum_p N\, c_p\,\underbrace{\raisebox{-.3cm}{\epsfysize=1.2cm\epsfbox{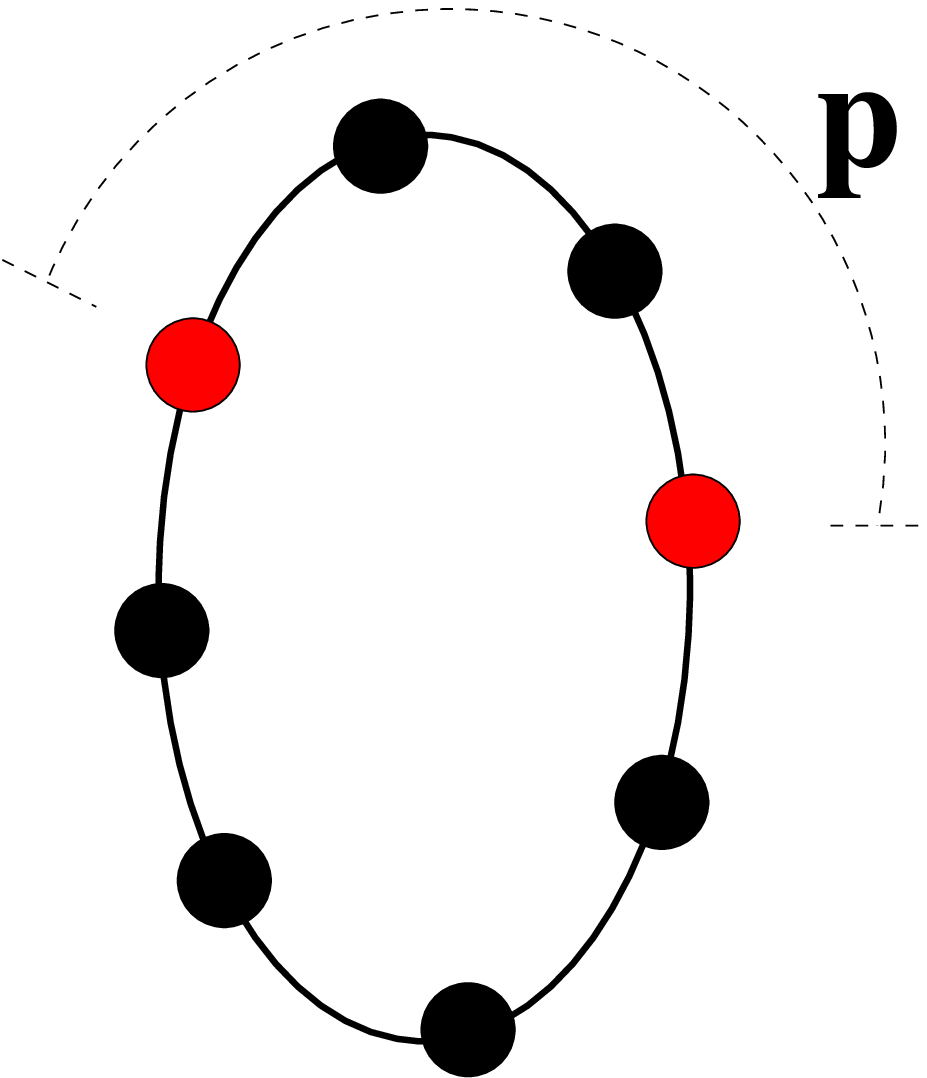}}}_{\rm planar}
+ \, \,d_p 
\underbrace{\raisebox{-.6cm}{\epsfysize=1.5cm\epsfbox{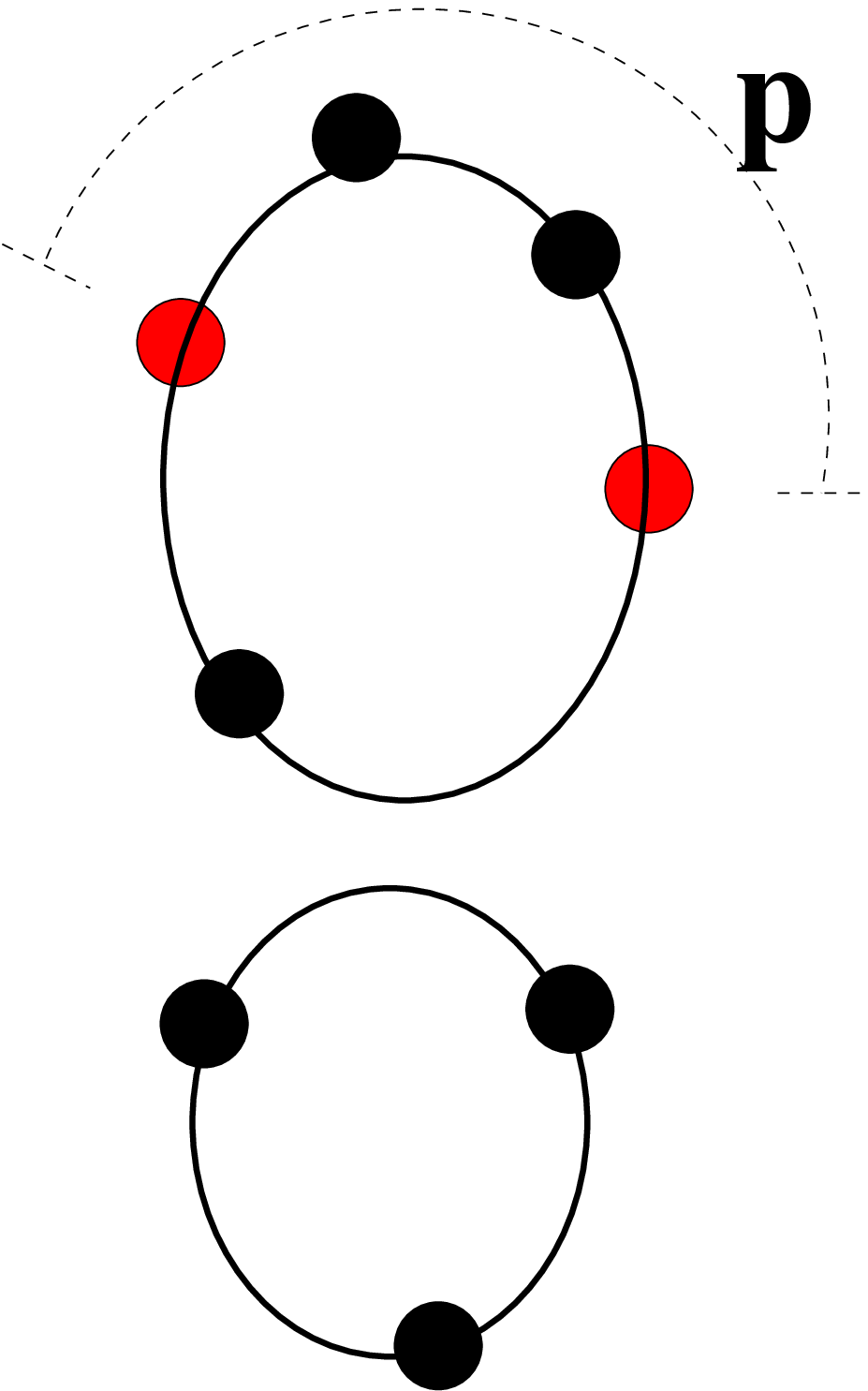}}}_{\rm nonplanar}\nn
$
\end{center}
\caption{The action of the dilatation operator on a trace operator.} 
\label{dilopact}
\end{figure}
``incoming" trace operators as depicted in figure~\ref{dilopact} and
may be computed in perturbation theory 
\begin{equation}
{\cal  D}
=\sum_{n=0}^\infty  {\cal D}^{(n)} \, ,
\end{equation}
where ${\cal D}^{(n)}$ is of order $\gym^{2n}$.
For the explicit computation of the one-loop piece ${\cal D}^{(1)}$ see e.g.~the
review \cite{hep-th/0307101}, where the concrete relation to 
two-point functions is also explained. In our `minimal' $SU(2)$ sector
of complex scalar fields $Z$ and $W$ it takes the rather simple form
\be
{\cal D}^{(0)} = \Tr(Z\check Z+W\check W),\qquad 
{\cal D}^{(1)}  =
-\frac{g^2_{\rm YM}}{8\pi^2}\, \Tr\, [Z,W]\, [\check{Z},{\check{W}}] \, ,
\qquad \mbox{where}\quad  \check{Z}_{ij}:=\frac{d}{dZ_{ji}} \, .
\label{D0n1}
\ee
Note that the tree-level piece ${\cal D}^{(0)}$ simply measures the length of the incident
operator (or spin chain) $J_1+J_2$.
The eigenvalues of the dilatation operator then yield the 
scaling dimensions we are looking for -- diagonalization of ${\cal D}$
solves the mixing problem.

We shall be exclusively interested in the planar contribution to ${\cal D}$, as 
this sector of the gauge theory corresponds to the ``free"
(in the sense of $g_s=0$) \AdSS string. For this it is important to realize that the
planar piece of ${\cal D}^{(1)}$ only acts on two neighboring fields
in the chain of $Z$'s and $W$'s. This may be seen by evaluating explicitly the action
of ${\cal D}^{(1)}$ on two fields $Z$ and $W$ separated by arbitrary matrices
$\cal{A}$ and $\cal{B}$
\be
\Tr\ [Z,W]\, [\check{Z},{\check{W}}]\circ
\Tr (Z\,{\cal A}\,  {W}\, {\cal B}) =
-\Tr ({\cal A})\, \Tr ( [Z,{W}]\, {\cal B})+ 
\Tr ({\cal B})\, \Tr ( [Z,{W}]\, {\cal A}) \, .
\ee
Clearly there is an enhanced contribution when ${\cal A}=1$ or ${\cal
  B}=1$, i.e.~$Z$ and $W$ are nearest neighbours on the spin chain.
From the above computation we learn that
\be
{\cal D}^{(1)}_{\rm planar} = \frac{\lambda}{8\pi^2}\, \sum_{i=1}^L\,
( \eins_{i,i+1} - P_{i,i+1} )
\ee
where $P_{i,j}$ permutes the fields (or
spins) at sites $i$ and $j$ and periodicity $P_{L,L+1}=P_{1,L}$ is understood. 
Remarkably, as noticed by Minahan and Zarembo
\cite{hep-th/0212208}, this spin-chain operator is the Heisenberg
XXX${}_{1/2}$ quantum spin chain Hamiltonian, which is the prototype of an 
integrable spin-chain. Written in terms of the Pauli matrices
$\vec\sigma_i$ acting on the spin at site $i$ one finds
\be
{\cal D}^{(1)}_{\rm planar}= \frac{\lambda}{8\pi^2}{\cal H}_{{\rm
    XXX}_{1/2}} = \frac{\lambda}{4\pi^2}\, \sum_{i=1}^L\,
(\frac{1}{4} -\vec\sigma_i\cdot\vec\sigma_{i+1}) \, .
\ee
Due to the positive sign in front of the sum, the spin chain is
ferromagnetic and its ground state is $|\down\down\ldots\down\,\rangle_{\rm cyclic}
\Leftrightarrow \Tr(Z^{L})$: The gauge dual of the rotating point particle of
section \ref{sect:particle}. Excitations of the ground state are
given by spin flips or ``magnons''. 
Note that a one-magnon excitation
$|\down\ldots\down\up\down\ldots \down\,\rangle_{\rm cyclic}$ has vanishing energy due
to the cyclic property of the trace, it corresponds to a zero mode plane-wave string
excitation $\alpha_0^{\dagger}\,|0\rangle$. Two-magnon excitations are the first stringy
excitations which are dual to the $\alpha_n^{\dagger}\alpha_{-n}^\dagger\, |0\rangle$
state of the plane-wave string in the BMN limit.

The integrability of the spin-chain amounts to the existence of $L-1$ higher charges $Q_k$ which
commute with the Hamiltonian (alias dilatation operator) and amongst themselves, 
i.e.~$[Q_k,Q_l]=0$. Explicitly the first two charges of the Heisenberg chain are given by
\be
Q_2:={\cal H}_{{\rm XXX}_{1/2}}\qquad
Q_3=\sum_{i=1}^L  (\vec\sigma_i\times \vec\sigma_{i+1})\cdot
\vec\sigma_{i+2} \, .
\ee
The explicit form of all the higher $Q_k$ may be found in \cite{hep-th/9403149}. Note that
$Q_k$ will involve up to $k$ neighboring spin interactions. 

\subsection{The coordinate Bethe ansatz}
\label{sect:coordBethe}

We now discuss the ansatz that enabled Bethe to diagonalize the Heisenberg model
in 1931 \cite{Bethe:1931hc}\footnote{For a nice and detailed review on this topic
see \cite{muller}. The technology of the
algebraic Bethe ansatz is reviewed in \cite{hep-th/9605187}.}. For this we will for 
the moment drop the cyclicity 
constraint imposed on us from the underlying trace structure of the gauge theory
operators and treat a general non-cyclic, but periodic, spin
chain. The vacuum state of the Heisenberg chain is then given by
$|\down\ldots\down\rangle$.
Let $|x_1,x_2,\ldots x_J\rangle$ with $x_1<x_2<\ldots <x_J$ 
denote a state of the chain with up-spins (magnons)
located at sites $x_i$, i.e.~ $|1,3,4\rangle_{L=5} =|\down\up\down\down\up\,\rangle$. 
It is useful to think of these spin flips as particles located at the sites $x_i$. Note that the
Hamiltonian preserves the magnon or particle number.

The one magnon sector is then trivially diagonalized by Fourier transformation
\begin{align}
|\psi(p_1)\rangle &:= \sum_{x=1}^L\, e^{i\, p_1\, x}\, |x\rangle ,\qquad
\mbox{with} \quad Q_2\, |\psi(p_1)\rangle = 4\, \sin^2(\frac{p_1}{2})
\, |\psi(p_1)\rangle\\
&\mbox{where} \quad Q_2= \sum_{i=1}^L\, ( \eins_{i,i+1} - P_{i,i+1} )
\end{align}
as $2-e^{ip}-e^{-ip}=4\,\sin^2(p/2)$. The periodic boundary condidtions
require the one-magnon momenta to be quantized $p_1=\frac{2\pi\,
  k}{L}$ with $k\in\Zint$.

Next consider a general two-magnon state of the form
\be
|\psi(p_1,p_2) \rangle =\sum_{1\leq x_1<x_2\leq L} \psi(x_1,x_2)\, |x_1,x_2\rangle\, .
\ee
with a two-particle wave-function $\psi(x_1,x_2)$. 
The ``position space'' Schr\"odinger equation following from 
$\sum_{i=1}^L(1-P_{i,i+1})\,|\psi(p_1,p_2)\rangle = {E_2}\, |\psi(p_1,p_2) \rangle$
then leads to two sets of equations, depending on whether the particles lie next to each other
or not:
\begin{align}
x_2&>x_1+1 \quad  &{E_2}\, \psi(x_1,x_2) =&\, 2\,\psi(x_1,x_2) -\psi(x_1-1,x_2)-\psi(x_1+1,x_2)\nn\\
              &   &                            +&\,2\,\psi(x_1,x_2) -\psi(x_1,x_2-1)-\psi(x_1,x_2+1)\nn\\
& &&\label{one}\\
x_2&=x_1+1   &{E_2}\, \psi(x_1,x_2) =& \,2\,\psi(x_1,x_2) -\psi(x_1-1,x_2)-\psi(x_1,x_2-1)\, .
\label{two}
\end{align}
$E_2$ is the eigenvalue of $Q_2$ and related to the gauge theory scaling dimensions
as $\Delta= L +\frac{\lambda}{8\pi^2}\, E_2 + {\cal O}(\lambda^2)$.
The above equations can be fulfilled by a superposition ansatz with an incoming and outgoing plane
wave (Bethe's ansatz)
\be
\label{Bethesansatz}
\psi(x_1,x_2)= e^{i({p_1}\, x_1 + {p_2}\, x_2)} + {S({p_2},{p_1})}\,
e^{i({p_2}\, x_1 + {p_1}\, x_2)} \,  ,
\ee
where $S(p_1,p_2)$ denotes the S-matrix of the scattered particles. Note that in the second
term describing the scattered contribution the two particles have simply exchanged their momenta.
One easily sees that
\eqn{one} is fulfilled for an arbitrary $S(p_2,p_1)$ yielding the energy as a sum of one-particle
energies
\be
E_2=4\, \sin^2 (\frac{p_1}{2}) + 4\, \sin^2 (\frac{p_2}{2}) \, .
\label{2ME}
\ee
Eq.~\eqn{two} then determines the S-matrix to be of the form
\be
S(p_1,p_2)=\frac{\varphi(p_1)-\varphi(p_2)+i}
{\varphi(p_1)-\varphi(p_2)-i}\qquad \mbox{with}\quad  \varphi(p)=\ft 1 2 \cot(\ft p 2)\, .
\label{S-matrix}
\ee
Note that $S(p_1,p_2)^{-1}=S(p_2,p_2)$.
This solves the infinite length chain. For a finite chain the momenta $p_i$ are no longer
arbitrary continuous quantities, but become discrete through the periodic boundary condition
\be
\psi(x_1,x_2)=\psi(x_2,x_1+L)\, .
\ee
This in turn then leads to the Bethe equations for the two magnon problem
\be
e^{ip_1 L}= S(p_1,p_2) \quad \mbox{and} \quad e^{ip_2 L}= S(p_2,p_1)
\label{2MBE}
\ee
implying $p_1+p_2=2\pi m$ with an arbitrary integer $m$. The solutions of the algebraic equations 
\eqn{2MBE} for $p_1$ and $p_2$
then determine the corresponding energies by plugging the resulting quasi-momenta $p_i$
into \eqn{2ME}.

The magic of integrability now is that this information is all that is needed to solve the general
$M$-body problem! This phenomenon is known as factorized scattering: The multi-body
scattering process factorizes into a sequence of two-body interactions under which
two incoming particles of momenta $p_i$ and $p_j$ scatter off each other {\it elastically}
with the S-matrix $S(p_j,p_i)$, thereby simply exchanging their momenta. 
That is the $M$-body wavefunction takes the form \cite{Bethe:1931hc,muller}
\be
\psi(x_1,\ldots ,x_M) = \sum_{P\in{\rm Perm}(M)} 
\exp\Bigl [ i\sum_{i=1}^M p_{P(i)}\, x_i  +\frac i 2 \sum_{i<j} \theta_{{P(i)P(j)}}\Bigr ]
\label{bethewf}
\ee
where the sum is over all $M!$ permutations of the labels
$\{1,2,\ldots, M\}$ and the phase shifts $\theta_{ij}=-\theta_{ji}$ are related to
the S-matrix \eqn{S-matrix} by
\be
S(p_i,p_j)=\exp[i\theta_{ij}]\, .
\ee
The $M$-magnon Bethe ansatz then yields the set of $M$ Bethe equations
\be
e^{ip_k L}=\prod_{i=1,i\neq k}^M S(p_k,p_i)
\label{BE}
\ee
with the two-body S-matrix of \eqn{S-matrix} and the additive energy expression 
\be
\label{E21loop}
E_2= \sum_{i=1}^M 4\, \sin^2 (\frac{p_i}{2})\, .
\ee
In order to reinstate the cyclicity of the trace condition one needs to further impose the 
constraint of a total vanishing momentum
\be
\sum_{i=1}^M p_i = 0\, .
\label{zerop}
\ee

As an example let us diagonalize the two magnon problem exactly. Due to \eqn{zerop} we have
$p:=p_1=-p_2$ and the Bethe equations \eqn{2MBE} reduce to the single equation
\be
e^{ipL} = \frac{\cot\frac{p}{2}+i}{\cot\frac{p}{2}-i} = e^{ip} \quad
\Rightarrow \quad e^{ip(L-1)}=1 \quad\Rightarrow \quad
p=\frac{2\pi\, n}{L-1} \, .
\ee
The energy eigenvalue then reads 
\be
E_2= 8 \sin^2\left (\frac{\pi\, n}{L-1}\right)\quad  \stackrel{L\to\infty}{\to}\quad
 8\pi^2\, \frac{n^2}{L^2} \, ,
\ee
which upon reinserting the dropped prefactor of $\frac{\lambda}{8\pi^2}$ yields
the one-loop scaling dimension $\Delta^{(1)}=\frac{\lambda}{\pi^2}\, \sin^2(\frac{\pi\, n^2\,}{L^2})$
of the two-magnon operators \cite{hep-th/0211032,hep-th/0212208}
\be
{\cal O}_n^{(J,2)} = \sum_{p=0}^J \cos\left( \pi\, n\, \frac{2p+1}{J+1}\right )\,
\Tr(W\, Z^p\, W\, Z^{J-p}) \, .
\ee
In the BMN limit $N,J\to\infty$ with $\lambda/J^2$ fixed the scaling dimension takes the
famous value $\Delta^{(1)}=n^2\,\lambda/J^2$, corresponding to the first term in the 
expansion of the level-two energy spectrum of the plane-wave superstring
$E_{\rm light-cone}=\sqrt{1+n^2\, \lambda/J^2}$ \cite{hep-th/0202021}.

Hence from the viewpoint of the spin chain the plane-wave limit corresponds to a 
chain of diverging length $L>>1$ carrying a finite number of magnons $M$, which are nothing
but the gauge duals of the oscillator excitations of the plane-wave superstring.

\subsection{The thermodynamic limit of the spin-chain}
\label{subsect:thermo}

In order to make contact with the spinning string solution discussed in section 
\ref{section:spinning_string_solutions} we will now consider the thermodynamic limit
of the spin chain in which the length $L$ {\it and} the number of magnons $M$
become large. This is necessary as the classical string solutions 
only limit to the true quantum result 
in the BMN type limit $J_2,J\to\infty$ with the filling fraction $J_2/J$ held fixed (here
$J_2=M$ and $J=L$). This thermodynamic groundstate solution of the gauge theory Bethe equations  
was found in \cite{hep-th/0306139,hep-th/0308117,hep-th/0310182,hep-th/0310188}
 which we closely follow.

For this it is useful to reexpress the Bethe equations \eqn{BE} in terms of the Bethe
roots $u_k$ related to the momenta via
\be
u_k=\frac{1}{2}\, \cot\frac{p_k}{2}
\ee
for which the Bethe equations \eqn{BE} and the momentum constraint \eqn{zerop} become
\be
\label{TBE}
\left ( \frac{u_i+i/2}{u_i-i/2}\right )^L = \prod_{k\neq i}^M 
\frac{u_i-u_k+i}{u_i-u_k-i}\, ,
\qquad \prod_{i=1}^M \frac{u_i+i/2}{u_i-i/2}= 1\, .
\ee
The energy then is
\be
\label{TE}
Q_2=\sum_{i=1}^M \frac{1}{{u_i}^2 + \ft 1 4}\, .
\ee
The momentum constraint can be satisfied by considering symmetric root distributions
of the form $(u_i,-u_i,u_i^\ast,-u_i^\ast)$.
The thermodynamic limit is now obtained by first taking the logarithm of \eqn{TBE}
\be
L\,\ln\left ( \frac{u_i+i/2}{u_i-i/2}\right ) 
= \sum_{k=1\, (k\neq i)}^M \ln \left ( \frac{u_i-u_k+i}{u_i-u_k-i} \right )
- 2\, \pi\, i\, n_i \, ,
\ee
where $n_i$ is an arbitrary integer associated to every root $u_i$. One
self-consistently assumes that the momenta scale as $p_i\sim 1/L$ for $L\to \infty$ 
implying that the Bethe roots scale as $u_i\sim L$.  Therefore in the $L\to\infty$
limit the above equation reduces to
\be
\label{BATL}
\frac{1}{u_i} = 2\pi\, n_j +\frac{2}{L} \sum_{k=1\, (k\neq i)}^M
\frac{1}{u_j-u_k} \, .
\ee
In the thermodynamic limit the roots $u_i$ accumulate on smooth contours in the
complex plane known as ``Bethe strings'' which turn the set of algebraic Bethe equations into
an integral equation. To see this introduce the Bethe root density 
\be
\label{RhoDef}
\rho(u):= \frac{1}{M} \sum_{j=1}^M \delta(u-u_j)
\qquad \mbox{with}\quad \int_C du\, \rho(u) =1 \, ,
\ee
where $C$ is the support of the density. i.e.~the union of all Bethe string contours. 
Multiplying \eqn{BATL} with
$u_i$ and introducing $\rho(u)$ one arrives at the singular integral 
equation \footnote{$\fpint{}{} dv\,
  \frac{(\ldots)}{v-u}$ denotes the principle part prescription.} 
 \be
\label{contBE}
\fpint{C}{} dv \, \frac{\rho(v)\, u}{v-u} = -\frac{1}{2\,\alpha} +
\frac{\pi\, n_{C(u)}\, u}{\alpha} \qquad \mbox{where}\quad  u\in C\,   \quad 
\mbox{and} \quad \alpha:=
\frac{M}{L}\, .
\ee
The mode numbers $n_{C(u)}$ are integers which are assumed to be
constant on each smooth component $C_n$ of the density support $C=\cup \, C_n$ in the complex
plane. These integers and the distribution of components $C_n$ select
the numerous solutions to the continuum Bethe equations \eqn{contBE}. Furthermore the 
continuum energy now becomes
\be
Q_2=M\, \int_C\, \frac{\rho(u)}{u^2} \, .
\ee

\begin{figure}[t]
\begin{center}
\includegraphics[width=9cm]{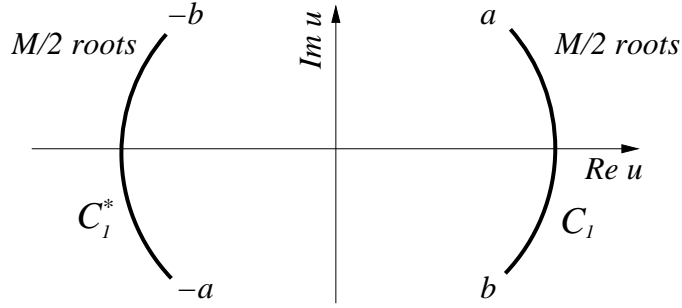}
\end{center}
\caption{Bethe root distribution for the gauge dual of the folded
  string. For large $L$ the roots condense into two cuts in the
  complex plane.}
 \label{roots}
\end{figure}

As was shown in \cite{hep-th/0306139} the gauge dual to the folded string 
solution of section~\ref{subsect:3.2} corresponds to a two cut 
support $C=C_1\cup C_1^\ast$ with $n_{C_1}=-1$ and
$n_{C_1^\ast}=1$ sketched in figure~\ref{roots}.
The key trick to obtain analytical expressions for $\rho(u)$ is to consider the 
analytic continuation to negative filling fraction $\beta:=-\alpha$: Then the two cuts
$C_1\cup C_1^\ast$ are mapped to intervals on the real line
($C^\ast_1\to [-b,-a]$ and $C_1\to [a,b]$) \cite{hep-th/0306139}.
Then \eqn{contBE} may be brought into the compact form 
\be
\label{FcontBE}
\fpint{a}{b} dv \, \frac{\tilde\rho(v)\, u^2}{v^2-u^2} = \frac{1}{4}-
\frac{\pi\, u}{2}
\qquad \mbox{with}\quad \int_a^b dv\, \tilde\rho(v)=\frac{\beta}{2}\, ,
\ee
using $\rho(-v)=\rho(v)$ and defining $\tilde\rho(v):=\beta\, \rho(v)$.
In order to proceed one introduces the resolvent
\be
H(u):=\int_a^b dv\, \tilde\rho(v)\, \frac{v^2}{v^2-u^2}=-\frac{\alpha}{2}+\sum_{k=1}^\infty
Q_{2k}\, u^{2k}
\ee
which gives rise to the infinite tower of conserved even charges $Q_{2k}$ 
with the energy $E_2=\ft{1}{8\pi^2}\,Q_2$ \cite{hep-th/0310182}. 
Across the cut $u\in[a,b]$ the resolvent $H(u)$ behaves as
\be
\label{Heqn}
H(u\pm i\, \epsilon) =-\frac{\alpha}{2} +\frac{1}{4} -\frac{\pi}{2}\, u
\pm i\,\pi \,\frac{u}{2}\, \tilde\rho(u)\, ,\qquad u\in[a,b]\,,
\ee
which one shows using the distributional identity $\frac{1}{x\pm i\, \epsilon}=
P(\frac{1}{x}) \mp i\,\pi \, \delta(x)$  and eq.~\eqn{FcontBE}.  From this
one obtains an integral expression for the resolvent
\be
\label{Hres}
H(u)=-\frac{\alpha}{2} +\frac{1}{4}-\int_a^b dv\, \frac{v^2}{v^2-u^2}\,
\sqrt{\frac{(b^2-u^2)\, (a^2-u^2)}{(b^2-v^2)\, (v^2-a^2)}}\, ,
\ee
which in turn self-consistently yields the density
\be
\tilde\rho(u)=\frac{2}{\pi\, u}\, \fpint{a}{b}  dv\, \frac{v^2}{v^2-u^2}\,
\sqrt{\frac{(b^2-u^2)\, (u^2-a^2)}{(b^2-v^2)\, (v^2-a^2)}}\, ,
\ee
Finally the interval boundaries $a$ and $b$ are implicitly determined through the normalization and
positivity conditions on $\tilde\rho(u)$ by the relations \cite{hep-th/0308117,hep-th/0310182}
\be
\frac{1}{a}=4\, K(q)\, ,\qquad \frac{1}{b}= 4\, \sqrt{1-q}\, K(q)\, ,
\qquad q:=\frac{b^2-a^2}{b^2}\, .
\ee
The resolvent and the density may be expressed in closed forms using the elliptic
integral of the third kind\footnote{Our convention is $\Pi(m^2,q):=\int_0^{\pi/2}
\frac{d\phi}{(1-m^2\,\sin^2\phi)\sqrt{1-q\,\sin^2\phi}}$\,.}
\begin{align}
H(u) &=-\frac{\alpha}{2}+\frac{1}{4}-\frac{\pi}{2}\, u -\frac{1}{4b}\,
\sqrt{\frac{a^2-u^2}{b^2-u^2}}\, \Bigl [ \frac{b^2}{a} - 4\, u^2\, \Pi
\Bigl(\frac{b^2-u^2}{b^2},
q\Bigr)\, \Bigr ]\, ,\nn\\
\tilde\rho(u) &=\frac{1}{2\,\pi\, b\, u}\,
\sqrt{\frac{u^2-a^2}{b^2-u^2}}\, \Bigl [ \frac{b^2}{a} - 4\, u^2\, \Pi
\Bigl(\frac{b^2-u^2}{b^2}, q\Bigr)\, \Bigr ]\, .
\end{align}
From this it is straightforward to (finally) extract the energy eigenvalue $Q_2$ of the
two cut solution in the parametric form
\be
\label{GTF1}
E_2= \frac{1}{2\pi^2}\, K(q)\, \Bigl [2\, E(q)-(2-q)\, K(q)\, \Bigr ]\,
\ee
with
\be
\label{GTF2}
\alpha=\frac{J_2}{J}= \frac{1}{2}-\frac{1}{2\,\sqrt{1-q}}\, \frac{E(q)}{K(q)}\, .
\ee
This final result for the one loop gauge theory anomalous scaling dimension can
now be compared to the folded string energies of section~\ref{subsect:3.2}
 eqs.~\eqn{FE1} and \eqn{FEprediction}.  They do not manifestly agree, however, if
 one relates the auxiliary parameters $q_0$ and $q$ through \cite{hep-th/0308117}
 \be
 q_0=-\frac{(1-\sqrt{1-q})^2}{4\, \sqrt{1-q}}
 \ee
 one may show that
 \be
 K(q_0)=(1-q)^{1/4}\, K(q) \, , \qquad E(q_0)=\ft 12 (1-q)^{-1/4}\, E(q) + \ft 12 (1-q)^{1/4}
 \, K(q)
 \ee
 using elliptic integral modular transformations. Using these relations the gauge theory
 result \eqn{GTF1} and \eqn{GTF2} may be transformed into the string result of
 \eqn{FE1} and \eqn{FEprediction}.  Hence the one-loop gauge theory scaling dimensions 
 indeed agree with the string prediction!

The analysis of the circular string configuration goes along the same lines. Here the Bethe roots 
turn out to condense on the imaginary axis. The root density is then symmetric along 
the imaginary axis, $\rho(u)=\rho(- u)$
and remains constant along a segment $[-c,c]$. For $u>c$ and $u<-c$ it falls off
towards zero. We shall not go through the detailed construction of  the density for this
configuration but refer the reader to the original papers of \cite{hep-th/0306139,hep-th/0308117}
being best explained in \cite{hep-th/0310182}. The outcome of this analysis is again a
perfect matching of the energy eigenvalue of the spin chain with 
the circular string energy of eqs.~\eqn{CS1} and \eqn{CS2}.

As a matter of fact one can go beyond this and match all the higher charges of gauge and string
theory, as was shown for the first time in \cite{hep-th/0310182} by using an approach based on the
B\"acklund transform. 

\subsection{Higher Loops in the SU(2) sector and discrepancies}

The discussed connection to an integrable spin chain at one-loop raises the question whether
integrability is merely an artifact of the one-loop approximation or a genuine property of planar
\Nfour gauge theory. Remarkably all present gauge theory data  points towards the 
latter being the case.

Higher loop contributions to the planar dilatation operator in the $SU(2)$ subsector
are by now firmly established to
the two-loop \cite{hep-th/0303060} and three-loop level \cite{hep-th/0310252,hep-th/0409009}. 
In  $s=1/2$ quantum spin chain language they take the explicit forms
\begin{align*}
 {\widehat{{\cal D}}_{\rm 2-loop}}  &= \sum_{i=1}^L
-\vec\sigma_l\cdot\vec\sigma_{l+2} + 4\,
\vec\sigma_l\cdot\vec\sigma_{l+1} - 3\cdot \eins \\
 {\widehat{{\cal D}}_{\rm 3-loop}}  &= \sum_{i=1}^L
-\vec\sigma_l\cdot\vec\sigma_{l+3} +
(\vec\sigma_l\cdot\vec\sigma_{l+2})\, (\vec\sigma_{l+1}\cdot\vec\sigma_{l+3})-
(\vec\sigma_l\cdot\vec\sigma_{l+3})\,
(\vec\sigma_{l+1}\cdot\vec\sigma_{l+2})\\& \qquad
+10 \,\vec\sigma_l\cdot\vec\sigma_{l+2} -29\,
\vec\sigma_l\cdot\vec\sigma_{l+1} +20\cdot \eins \, .
\end{align*}
In general the $k$-loop contribution to the dilatation operator involves interactions of $k+1$
neighboring spins, i.e.~the full dilatation operator ${\cal D}=\sum_{k=1}^\infty
{\cal D}_{\rm k-loop}$ will correspond to a long-range interacting spin-chain
hamiltonian. Note also the appearance of novel quartic spin interactions $(\vec\sigma_i\cdot\vec\sigma_j)
(\vec\sigma_k\cdot\vec\sigma_l)$ at the three-loop level. Generically even higher interactions
of the form $(\vec\sigma_\bullet\cdot\vec\sigma_\bullet)^k$ are expected at the $[\frac k 2]+1$ loop
levels.
 
Integrability remains stable up to the three-loop order and acts in a 
perturbative sense:
The conserved charges of the Heisenberg XXX${}_{1/2}$ chain receive higher order corrections  in 
$\lambda$ of the form
$
{\cal Q}_k= Q_k^{(1)}+\lambda\, Q_k^{(2)} +\lambda^2\, Q_k^{(3)}+\ldots
$ as one would expect.
The full charges ${\cal O}_k$ commute with each other ($[{\cal O}_k,{\cal O}_l]=0$) 
which translates into commutation relations for the various loop contributions
${\cal O}^{(r)}_k$ upon expansion in $\lambda$,
i.e.
\begin{align}
[Q^{(1)}_k,Q^{(1)}_l]=0\, ,\qquad [Q^{(1)}_k,Q^{(2)}_l]+[Q^{(2)}_k,Q^{(1)}_l]&=0\, \nonumber\\
[Q^{(1)}_k,Q^{(3)}_l]+[{Q}^{(2)}_k,{Q}^{(2)}_l]+[Q^{(3)}_k,Q^{(1)}_l]&=0
\end{align}
and so on. However, opposed to the situation for the Heisenberg chain
\cite{hep-th/9605187}, 
there does not yet exist
an algebraic construction of the gauge theory charges at higher loops. Nevertheless the
first few $Q_k$ have been constructed manually to higher loop-orders \cite{hep-th/0409054}. 

An additional key property of these higher-loop corrections is that they obey BMN
scaling: The emergence of the effective loop-counting parameter $\lambda':=\lambda/J^2$
in the $J\to\infty$ limit  leading to the scaling dimensions $\Delta\sim\sqrt{1+\lambda'\, n^2}$
for two magnon states in quantitative agreement with plane-wave superstrings.

Motivated by these findings Beisert, Dippel and Staudacher \cite{hep-th/0405001} 
turned the logic around and simply {\sl assumed} integrability, BMN scaling and a Feynman
diagrammatic origin of the $k$-loop $SU(2)$ dilatation operator. Interestingly these assumptions
constrain the possible structures of the planar dilatation operator completely up to
the five-loop level (and possibly beyond).

How can one now diagonalize the higher-loop corrected dilatation operator? For this the
ansatz for the Bethe wave-functions \eqn{Bethesansatz} needs to be adjusted in a perturbative 
sense in order to accommodate the long-range interactions, leading to a 
``perturbative asymptotic Bethe ansatz'' for the two magnon wave function \cite{hep-th/0412188} 
\begin{align}
\psi(x_1,x_2)&= e^{i({p_1}\, x_1 + {p_2}\, x_2)}\, {f(x_2-x_1,p_1,p_2)}\nn\\
& \quad + {S(p_2,p_1)}\,
e^{i({p_2}\, x_1 + {p_1}\, x_2)}\, {f(L-x_2+x_1,p_1,p_2)} \, .
\label{PABA}
\end{align}
Here one needs to introduce a perturbative deformation of the S-matrix
\be
\label{PABAS}
S(p_1,p_2)=S_0(p_1,p_2)+\sum_{n=1}^\infty {\lambda^n}\, S_n(p_1,p_2)
\ee
which is determined by the eigenvalue problem. Moreover suitable  ``fudge functions'' 
enter the ansatz
\be
{f(x,p_1,p_2)}=1+ \sum_{n=0}^\infty {\lambda^{n+|x|}}\, f_n(x,p_1,p_2)
\stackrel{x\gg 1}{\longrightarrow} 1 \,
\ee
which account for a deformation of the plane-wave form of the eigenfunction when two magnons
approach each other within the interaction range of the spin-chain hamiltonian\footnote{We are here actually
using a slightly modified definition of these functions to the one presented in \cite{hep-th/0412188}
which was considered in \cite{hep-th/0412331}.}. By construction they are invisible in the 
asymptotic regime $x\gg 1$ (or rather $x$ larger than the highest loop order considered) of well separated
magnons. The detailed form of these functions is completely irrelevant for the physical spectrum as
a consequence of the factorized scattering property of the integrable system.

With this perturbative asymptotic Bethe ansatz \eqn{PABA} one shows that the form of the
Bethe equations remains unchanged, i.e.~the perturbative S-matrix \eqn{PABAS} simply
appears on the right hand side of the equations
\be
e^{ip_k L}=\prod_{i=1,i\neq k}^M S(p_k,p_i) \, ,
\ee
and is determined by demanding $\psi(x_1,x_2)$ to be an eigenfunction of the dilatation operator
just as we did in section~\ref{sect:coordBethe}.
Based on the constructed five-loop form of the dilatation operator the S-matrix is then determined up
to ${\cal O}(\lambda^4)$. The obtained series in $\lambda$ turns out to be of a remarkably simple
structure, which enabled the authors of \cite{hep-th/0405001} to conjecture 
an asymptotic {\sl all} loop  expression for the perturbative S-matrix 
 \be
 \label{BDS1}
S(p_1,p_2)=\frac{\varphi(p_1)-\varphi(p_2)+i}{\varphi(p_1)-\varphi(p_2)-i}
\qquad \mbox{with} \quad
\varphi(p)= \ft 1 2\, \cot (\ft p 2)\, \sqrt{1+{\lambda}\, \sin^2(\ft p 2)} \, ,
\ee
to be compared to the one-loop form of \eqn{S-matrix}.  The conjectured asymptotic all loop form of the
energy density generalizing the one-loop expression \eqn{E21loop} reads\footnote{The full
conjecture for all the higher charge densities $q_k$ may be found in \cite{hep-th/0405001}.}
\be
\label{BDS2}
\lambda\, q_2(p)=  \sqrt{1+8\,\lambda\,\sin^2(\ft p2)}-1 
\ee
with the total energy being given by $E_2=\sum_{i=1}^M q_2(p_i)$. Note that both expressions
manifestly obey BMN scaling as the quasi-momenta scale like $p\sim L^{-1}$ in the thermodynamic limit 
as we discussed in section~\ref{subsect:thermo}. It is important to stress that these Bethe equations
only make sense asymptotically: For a chain (or gauge theory operator) of
length $L$ the eqs.~\eqn{BDS1} and \eqn{BDS2} yield a prediction for the energy (or
scaling dimension) up to $L-1$ loops. This is so, as the interaction range of the Hamiltonian will
reach the length of the spin chain beyond this point, and the multi-magnon wavefunctions of \eqn{PABA} 
can never enter the asymptotic regime. What happens beyond the $L$ loop level is still a mystery.
At this point the ``wrapping" interactions 
start to set in: The interaction range of the spin chain
Hamiltonian cannot spread any further and starts to ``wrap" around the chain. In the dimensional
reduced model of plane-wave matrix theory 
\cite{hep-th/0202021,hep-th/0205185,hep-th/0207034,hep-th/0207050,hep-th/0306054,hep-th/0310232}
these effects have been studied
explicitly at the four-loop level in \cite{hep-th/0412331} where the wrapping effects set
in for the first time in the $SU(2)$ subsector. No ``natural" way of transforming the generic 
dilatation operator to the wrapping situation was found. Finally let us restate that it has not been shown 
so far that a microscopic long-range spin-chain Hamiltonian truly exists, which has a 
spectrum determined by the conjectured perturbative asymptotic Bethe equations \eqn{BDS1} 
and \eqn{BDS2} of Beisert, Dippel and Staudacher.

\medskip

In any case, the proposed all-loop asymptotic Bethe equations \eqn{BDS1} and \eqn{BDS2} may now
be studied in the thermodynamic limit just as we did above for the one-loop case. This
was done in \cite{hep-th/0401057} and \cite{hep-th/0405001} and allows
for a comparison to the results obtained in section~\ref{section:spinning_string_solutions}
for the energies of the spinning folded and closed string solutions. Recall that these yield
predictions to all-loops in $\lambda'$. While the two loop gauge theory result is in perfect
agreement, the three-loop scaling dimensions {\sl fail} to match with the expected dual 
string theory result! This three-loop disagreement also arises in the comparison to the 
near plane-wave string spectrum computed in 
\cite{hep-th/0307032,hep-th/0404007,hep-th/0405153,hep-th/0407096}, i.e.~the first
$1/J$ corrections to the Penrose limit of \AdSS to the plane-wave background.

Does this mean that the AdS/CFT correspondence does not hold in its
strong sense? While this logical possibility certainly
exists,  an alternative explanation is that one is dealing with an
order of 
limits problem as pointed out 
initially in \cite{hep-th/0405001}. While in string theory one works
in 
a limit of $\lambda\to\infty$
with $J^2/\lambda$ held fixed, in gauge theory one stays in the perturbative regime $\lambda\ll 1$
and thereafter takes the $J\to\infty$ limit, keeping only terms which scale as $\lambda/J^2$.
These two limits need not commute. Most likely the above-mentioned ``wrapping" interactions
must be included into the gauge theory constructions in order to match the string theory energies.
On the other hand, the firm finite $L$ results at order $\lambda^3$ of
the gauge theory are still free of ``wrapping"
interactions: These only start to set in at the four-loop level (in the considered minimal $SU(2)$
subsector). Moreover to what extent the integrability
is preserved in the presence of these ``wrapping" interactions is unclear at the moment.
Certainly the resolution of this discrepancy remains a pressing open problem in the field.

\subsection{Further developments}

The gauge theory analysis of the planar dilatation operator and its relation to
integrable spin chain models has been extended in two directions: Larger sectors within \Nfour super
Yang-Mills and conformal deformations of the original theory. 

In their seminal paper uncovering the integrable spin chain structure Minahan and
Zarembo \cite{hep-th/0212208} actually considered the full scalar sector of 
the gauge theory at one-loop order. This gives rise to an integrable $SO(6)$ magnetic quantum spin chain
of which the discussed $SU(2)$ Heisenberg model arises in a subsector.
This work was generalized in \cite{hep-th/0307015} 
to all local operators of the planar one-loop \Nfour theory, leading to an integrable super-spin
chain with $SU(2,2|4)$ symmetry discussed in \cite{hep-th/0307042}. 
The excitations of this super-spin chain consist of scalars, field strengths, fermions and an
arbitrary number of covariant derivatives of these three leading to an infinite number of spin
degrees of freedom on a single lattice site. 
The thermodynamic limit of this 
super spin chain was later on constructed in larger supersymmetric subsectors in \cite{hep-th/0412254} and 
in \cite{hep-th/0503200} for the full system leading 
to spectral curves, which reproduce the results of the classical string at one-loop order.

The conjectured form of the asymptotic higher-loop Bethe ansatz for the $PSU(2)$ subsector was
generalized to the full theory recently  in \cite{hep-th/0504190} in
form of a long-range $SU(2,2|4)$
Bethe ansatz. In this paper the corresponding generalization for the
quantum string Bethe equations,
generalizing \cite{hep-th/0406256} relevant for the $SU(2)$ sector, was also provided. 
A novel feature of leaving the minimal $SU(2)$ sector at higher loop orders
is that the spin chain begins to fluctuate in length \cite{hep-th/0310252}: The hamiltonian 
(or dilatation operator) preserves
the classical scaling dimensions but not the length of the chain,
e.g.~two fermions have the same classical
scaling dimension as three scalars and can mix if they carry identical
charges.

An alternative route for comparing string energies to gauge theory scaling dimensions lies in the
coherent-state effective action approach pioneered by Kruczenski \cite{hep-th/0311203}.
Here one establishes an effective action for the string whose center of mass moves along a big
circle of the $S^3$ with large angular momentum in the ``weak
coupling" limit $\lambda/J^2\ll 1$.
This action is then shown to agree with the long-wave length
approximation of the gauge theory spin chain at one-loop. In this
approach there is no need to compare explicit solutions any longer,
however, considering higher-loop effects and fermions becomes more challenging in this language.
For details see
\cite{hep-th/0403120,hep-th/0403139,hep-th/0404133,hep-th/0406189,hep-th/0410022,hep-th/0503185,hep-th/0409086,hep-th/0503159}
and also Tseytlin's review \cite{hep-th/0409296}.

In view of the reviewed insights an obvious next question to address is what can 
be said about
the non-planar sector of the gauge theory dual to string interactions. The gauge theory dilatation
operator is indeed known for the first two loop orders in the $SU(2)$ sector exactly, that is
including all non-planar contributions. However, extracting
physical data from it, such as amplitudes for the decay of single trace operators to double 
trace ones is nontrivial. This has been performed with great success for the case of two or three
magnon excitations in the BMN limit being dual to the interacting plane-wave superstring  
(reviewed in \cite{hep-th/0307027}). Performing the same computation for a macroscopic
number of magnons, thus describing the quantum decay of the discussed spinning strings,
is complicated enormously by the complexity of the Bethe
wavefunction \eqn{bethewf} for large $M$. The analysis on the string side, however, can be
performed by considering a semiclassical decay process \cite{hep-th/0410275}. 
For a related discussion on the non-planar gauge theory aspects see \cite{hep-th/0404066}.

An interesting toy model for \Nfour Super Yang-Mills is its dimensional reduction on
a three sphere to a quantum mechanical system  \cite{hep-th/0306054}, which turns out to be the 
plane-wave matrix theory of
\cite{hep-th/0202021,hep-th/0205185,hep-th/0207034,hep-th/0207050} related to M-theory on
the plane-wave background. 
The Hamiltonian of this matrix quantum mechanics
reduces to an integrable spin chain in the large $N$ limit, which remarkably is identical to 
the full \Nfour system up to the three loop level in the overlapping $SU(2|3)$ sector 
\cite{hep-th/0310232} (via a perturbative 
redefinition of the coupling constant). However, a recent four loop study displays a breakdown
of BMN scaling at this level of perturbation theory while integrability is stable 
\cite{hep-th/0412331}. 
What this finding implies for the full \Nfour model remains to be seen.

A recent line of research concerns the study of deformations of ${\cal
  N}=4$ Super Yang-Mills
which maintain the quantum conformal structure known as the Leigh-Strassler or $\beta$
deformations \cite{hep-th/9503121}. The one-loop dilatation operator was constructed in
subsectors of the theory in \cite{hep-th/0312218,hep-th/0405215}. Moreover the explicit
construction of the supergravity background dual to the $\beta$ deformed gauge theory
was achieved by Lunin and Maldacena \cite{hep-th/0502086}. Again the classical bosonic
string theory in this background is integrable and exhibits a Lax pair \cite{hep-th/0503192}
yielding a string Bethe equation which agrees with the thermodynamic limit of the one-loop
Bethe equation for the gauge theory dilatation operator \cite{hep-th/0503201}. So the complete
discussion of this review lifts to the $\beta$ deformed case. As a matter of fact even larger
(three-parameter families) of generically non-supersymmetric deformations of $\Nfour$ 
Super Yang-Mills have been considered with known supergravity duals \cite{hep-th/0503201}.
The corresponding ``twisted" gauge theory spin chain and Bethe ansatz was constructed in
\cite{hep-th/0505187}. 

Open integrable spin chains have also appeared in the AdS/CFT setting where the
boundaries of the spin chain correspond to fields in
the fundamental representation, see 
\cite{hep-th/0312091,hep-th/0401016,hep-th/0403004,hep-th/0401041} for
such constructions. In \cite{hep-th/0501078} open
integrable spin chains emerged within subdeterminant operators in \Nfour
super Yang-Mills dual to so-called ``giant gravitons''.

First investigations on the role of integrability for the three-point functions in the gauge
theory were performed in \cite{hep-th/0404190,hep-th/0407140,hep-th/0502186}. 

Finally let us mention that integrable structures are known to also
appear in QCD, such as in high-energy scattering processes and other
instances \cite{hep-th/9311037,hep-th/9404173,hep-ph/9805225,hep-ph/9907420},
see \cite{hep-th/0407232} for a recent review.

In conclusion the emergence of integrable spin chains in the AdS/CFT correspondence has led
to great insights into dynamical aspects of the duality and might hold the key to a complete
determination of the spectrum of both theories. Recent developments point towards 
integrability being a generic 
property of conformal gauge theories in the planar limit not necessarily connected to supersymmetry. 
Finally a great challenge for the future is to 
understand the integrable spin chain nature of the {\sl quantum} string in \AdSS and related backgrounds.

\newpage


\section{Acknowledgements}

I wish to thank Gleb Arutyunov, Niklas Beisert, Sergey Frolov,
Matthias Staudacher, Arkady Tseytlin and Marija Zamaklar for helpful discussions and
important comments on the manuscript. This review grew out of a lecture
delivered at the Post-Strings 2004 Meeting at Durham. I 
thank the organizers of the meeting for hospitality and an inspiring workshop.

\newpage



\bibliography{lrl}

\end{document}